\documentclass[11pt,letterpaper]{article}
\usepackage{soul}
\usepackage{jheppub}
\usepackage[utf8]{inputenc}
\usepackage[T1]{fontenc}
\usepackage{amsfonts}
\usepackage{tikz}
\usepackage{hyperref}
\usepackage{comment}
\usepackage{amsmath}
\usepackage{graphicx,subfigure}
\usepackage{tensor} 
\usepackage{mathtools} 
\usepackage{amssymb}
\usepackage{enumerate}
\usepackage{euscript}
\usepackage{indentfirst}
\usepackage{color}
\usepackage{xcolor}
\usepackage{float}
\usepackage{braket}
\usepackage[normalem]{ulem}
\usepackage{cancel}

\newcommand{\tr}{\mathrm{Tr}}
\newcommand{\nn}{\nonumber}
\newcommand{\ads}{{\rm AdS}}
\newcommand{\lads}{\ell_{\rm AdS}}
\newcommand{\ds}{{\rm dS}}
\newcommand{\lds}{\ell_{\rm dS}}
\newcommand{\eff}{{\rm eff}}
\newcommand{\ladsds}{\ell_{(\rm{A})dS}}

\title{ \boldmath Holographic $T\bar{T}$ deformation of the entanglement entropy in (A)dS$_3$/CFT$_2$}

\author[]{Jing-Cheng Chang$^{a,}$$^{b,}$$^c$,}
\author[]{Song He$^{d,}$$^{e,}$$^{f,}$$^g$,}
\author[]{Yu-Xiao Liu$^{a,}$$^{b,}$$^c$ and}
\author[]{Long Zhao$^f$}

\emailAdd{120220908561@lzu.edu.cn, hesong@jlu.edu.cn, liuyx@lzu.edu.cn, zhaolong@jlu.edu.cn}

\affiliation[a]{Key Laboratory for Quantum Theory and Applications of MoE, Institute of Theoretical Physics and Research Center of Gravitation, Lanzhou University, Lanzhou 730000, China
	\vspace{0.1cm}}

\affiliation[b]{Key Laboratory of Theoretical Physics of Gansu Province, Lanzhou Center for Theoretical Physics,  Lanzhou University, Lanzhou 730000, China 
	\vspace{0.1cm}}

\affiliation[c]{School of Physical Science and Technology,
	Lanzhou University, Lanzhou 730000, China 
	\vspace{0.1cm}}

\affiliation[d]{Institute of Fundamental Physics and Quantum Technology,  \newline Ningbo University, Ningbo, Zhejiang 315211, China 
	\vspace{0.1cm}}

\affiliation[e]{School of Physical Science and Technology, 
	Ningbo University, Ningbo, 315211, China
	\vspace{0.1cm}}

\affiliation[f]{Center for Theoretical Physics and College of Physics,
	\newline Jilin University, Changchun 130012, People's Republic of China
	\vspace{0.1cm}}

\affiliation[g]{Max Planck Institute for Gravitational Physics (Albert Einstein Institute),  
	\newline Am Mühlenberg 1, 14476 Golm, Germany}

\abstract{In recent years, the holographic duality between $T\bar{T}$-deformed conformal field theory (CFT) and Anti-de Sitter (AdS) spacetime with finite radial cutoff has received significant attention. The study of $T\bar{T}$ deformation within the framework of de Sitter (dS)/CFT duality has also progressed. This paper shows that the trace flow equation in dS spacetime can be analytically extended from its AdS counterpart through double Wick rotations. Meanwhile, we generalize the replica method in both AdS and dS holography to derive a general expression for the entanglement entropy of arbitrary single spatial intervals within the $T\bar{T}$-deformed framework. For both finite size and finite temperature systems, we obtain the analytical expression for the entanglement entropy after $T\bar{T}$ deformation. Finally, in dS/dS holography and half-dS holography, we find that the dual field theory exhibits non-locality by analyzing the strong subadditivity and boosted strong subadditivity of entanglement entropy.}


\begin{document}
	\maketitle
	\flushbottom
	
	
	\section{Introduction}
\begin{figure}[H]
		\centering
		\includegraphics[scale=0.8]{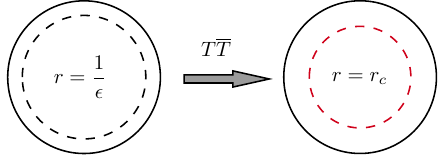}
		\caption{The holography $T\bar{T}$ deformation. The black solid circle represents the AdS boundary, the black dashed circle represents the cutoff at the asymptotic boundary $r=\frac{1}{\epsilon}$, and the red dashed circle represents the finite cutoff $r=r_c$.}
		\label{tt}
	\end{figure}

	\noindent 
	The $T\bar{T}$ deformation~\cite{Smirnov:2016lqw, Cavaglia:2016oda} of conformal field theory (CFT) is an irrelevant deformation that exhibits factorization properties in the large central charge limit~\cite{Zamolodchikov:2004ce, Jiang:2019tcq}. For a two-dimensional CFT, the $T\bar{T}$ deformation is defined by the double trace operator
	\begin{equation}
		\frac{dS}{d\lambda}=2\pi\int d^2x\sqrt{g}~T\bar{T}, \quad T\bar{T}\equiv\frac18(T^{ij}T_{ij}-(T_i^i)^2).
	\end{equation}
	The strength of this deformation is controlled by the deformation parameter $\lambda$. If we treat the deformation as a perturbation and expand the deformed theory to first-order in the deformation parameter $\lambda$, we obtain 
	\begin{equation}
		S_{\mathrm{QFT}}\simeq S_{\mathrm{CFT}}+2\pi\lambda\int d^2x\sqrt{g}~T\bar{T}.
		\label{eq:action-ttbar-order1}
	\end{equation}
	From the perspective of the AdS/CFT duality~\cite{Maldacena:1997re, Gubser:1998bc, Witten:1998qj}, the $T\bar{T}$-deformed CFT corresponds to the semiclassical gravity theory living in an AdS spacetime with finite cutoff~\cite{McGough:2016lol, Kraus:2018xrn}, as shown in figure~\ref{tt}.
	
	In refs.~\cite{McGough:2016lol,Kraus:2018xrn}, the authors calculated the propagation speed of the massless modes, energy spectrum, and partition functions in both the contexts of the $T\bar{T}$-deformed quantum field theory and AdS gravity with a finite cutoff. The generic holographic correlation functions have been established in the refs.~\cite{He:2023hoj, He:2023knl, He:2024fdm, He:2024xbi}. These results support the duality depicted in figure~\ref{tt}. In addition to the aforementioned correspondences, another significant result of holographic principles is that the entanglement entropy of the boundary field theory equals the area of the minimal surface, also known as the Ryu-Takayanagi (RT) surface, inside the AdS spacetime~\cite{Hubeny:2007xt, Ryu:2006bv}. The RT formula for holographic entanglement entropy serves as a criterion for validating the holographic principle, with $T\bar{T}$-deformed theories offering an additional test. For the holographic duality with $T\bar{T}$ deformation, this duality has been discussed in refs.~\cite{Chen:2018eqk, Jeong:2019ylz, Jiang:2019epa,Allameh:2021moy, Setare:2022qls, He:2023xnb, Apolo:2023ckr, Apolo:2023vnm}. 
	Meanwhile, the reflected entropy~\cite{Dutta:2019gen} and pseudo entropy~\cite{Nakata:2020luh} of the $T\bar{T}$-deformed field theory have been studied in refs.~\cite{Basu:2024bal, He:2023wko}. For more interesting work on $T\bar{T}$ deformation, please refer to refs.~\cite{He:2019vzf, Asrat:2020uib,Khoeini-Moghaddam:2020ymm, Li:2020pwa, Jeong:2022jmp, He:2023obo, Tian:2023fgf,Basu:2023bov,Basu:2023aqz, He:2024pbp,Tsolakidis:2024wut, Babaei-Aghbolagh:2024hti,He:2025ppz,He:2025fdz}.
	
 Although the holographic principle has been well established in the AdS spacetime with finite cutoff $r=r_c$, it remains crucial to investigate whether the holographic principle extends to dS spacetime~\cite{Witten:2001kn, Maldacena:2002vr, Anninos:2012ft, Miyaji:2015yva, Narayan:2017xca, Hikida:2021ese,Chen:2022ozy, Chang:2023gkt,Chen:2023prz,Doi:2024nty}, since our universe is more accurately described by a dS metric. In the standard dS/CFT duality~\cite{Strominger:2001pn,Strominger:2001gp}, the CFT resides at a future spacelike infinite, which leads to some exotic characteristics~\cite{Anninos:2011ui,Ng:2012xp, Hikida:2022ltr, Doi:2022iyj}. For example, the central charge is imaginary. Based on these exotic characteristics, a precise construction of the dual CFT in dS spacetime remains elusive. For the holographic entanglement entropy in dS/CFT, by calculating the extreme value of the area of the RT surface anchored at the future spacelike infinite boundary on the gravitational side, refs.~\cite{Narayan:2015vda,Narayan:2015oka,Narayan:2016xwq, Narayan:2022afv,Narayan:2023zen} have revealed the presence of complex-valued saddle points, which coincides with the imaginary central charge. At the same time, this observation aligns with results obtained from the analytic continuation of holographic entanglement entropy in Euclidean AdS spacetime, thereby offering a potential avenue to explore the dS/CFT duality through double Wick rotation~\cite{Sato:2015tta}. In addition, other holographic duality models in the dS spacetime, such as the dS/dS holography~\cite{ Alishahiha:2004md,  Alishahiha:2005dj, Dong:2018cuv,Geng:2019bnn, Geng:2019ruz,Geng:2020kxh, Geng:2021wcq} and half-dS holography~\cite{Kawamoto:2023nki}, have also gained attention and been studied. In these two models, the dual field theory resides on the timelike boundary of the dS spacetime.
		
		The primary purpose of this paper is to study the impact of the $T\bar{T}$ deformation on holographic entanglement entropy in the dS spacetime, including three holographic models: dS/CFT, dS/dS, and half-dS. For the $T\bar{T}$-deformed dS/CFT~\cite{Chen:2023eic}, the dual field theory is on the spacelike slice $t=t_c$, which coincides with the idea of Cauchy slice holography~\cite{Araujo-Regado:2022gvw, Araujo-Regado:2022jpj}. In the $T\bar{T}$-deformed dS/dS, the cosmological constant term also appears in the trace flow equation of the $T\bar{T}$ deformation, thus the $T\bar{T}$ deformation of this model is referred to as $T\bar{T}+\Lambda_2$ deformation~\cite{Gorbenko:2018oov,Shyam:2021ciy,Coleman:2021nor,Torroba:2022jrk,Batra:2024kjl}. However, the deformed field theory no longer retains conformal symmetry, rendering the original method for calculating the entanglement entropy inapplicable. Fortunately, when the $T\bar{T}$-deformed CFT is situated on a two-dimensional sphere, the entanglement entropy of a subregion with endpoints at antipodal points can be calculated by replicating the spherical geometry~\cite{Calabrese:2009qy,Donnelly:2018bef, Caputa:2019pam, Deng:2023pjs, Grieninger:2019zts}. In this context, by adjusting the spherical radius $r$, we can also study the renormalization group flow~\cite{Wilson:1973jj, Polchinski:1983gv} of the dual field theory under $T\bar{T}$ deformation, with $r$ functioning as the radial coordinate in the bulk spacetime. This method is an extension of the approach proposed in ref.~\cite{Lewkowycz:2013nqa} and was further generalized in ref.~\cite{Lewkowycz:2019xse} to compute the entanglement entropy of arbitrary spatial intervals on a two-dimensional plane. In ref.~\cite{Chen:2023eic}, the replica method was employed within the $T\bar{T}$-deformed dS/CFT duality to calculate the entanglement entropy for the subregion between two antipodal points on the sphere. In the $T\bar{T}$-deformed theories, non-perturbative calculations of entanglement entropy have been achieved using the replica trick, yielding exact results on spherical and flat manifolds~\cite{Donnelly:2018bef, Lewkowycz:2019xse}. However, the non-perturbative calculation on general manifolds is still lacking. 
	
	In this paper, we present a general expression for the entanglement entropy of arbitrary spatial intervals in general manifolds by combining the Casini-Huerta-Myers (CHM) map~\cite{Casini:2011kv} with the replica method on a sphere. The results are	
	\begin{itemize}
		\item AdS($0<r_{\text{eff}}<\infty$):
		$S=\frac{c_\ads}{3}\tanh^{-1}\left(\frac{1}{\sqrt{1+\frac{c_\ads\lambda_\ads }{12r_{\text{eff}}^2}}}\right),$
		\item dS($\lds <r_{\text{eff}}<\infty$):
		$S=-\frac{ic_\ds}{3}\coth^{-1}\left(\frac{1}{\sqrt{1-\frac{c_\ds\lambda_\ds }{12r_{\text{eff}}^2}}}\right)+\frac{\pi c_\ds}{6},$
		\item dS($0<r_{\text{eff}}<\lds $): 
		$S=\frac{c_\ds}{3}\tan^{-1}\left(\frac{1}{\sqrt{\frac{c_\ds\lambda_\ds }{12r_{\text{eff}}^2}-1}}\right),$
	\end{itemize}
	where $\ladsds$ represents the (A)dS radius, $c_{\rm(A)dS}=\frac{3\ladsds}{2G}$ is the Brown-Henneaux central charge~\cite{Brown1986},  $\lambda_{\rm(A)dS}=8G\ladsds$ is the $T\bar{T}$ deformation parameter\footnote{As mentioned in ref.~\cite{Donnelly:2018bef}, we adopt the same unit system. The factor $r_c$, which distinguishes the CFT metric from the induced bulk metric, is absorbed into the effective radius $r_{\text{eff}}$ of the sphere.} and \(r_{\text{eff}}\) denotes the effective radius of the sphere, which is related to the size of the subsystem. In the case of dS holography, we find that the results for the holographic entanglement entropy can be summarized in a single formula:
	\begin{equation}
		S=-\frac{ic_\ds}{3}\tanh^{-1}\left(\frac{1}{\sqrt{1-\frac{c_\ds\lambda_\ds }{12r_{\text{eff}}^2}}}\right)\,.
	\end{equation}
	When the effective radius of the sphere is greater than the cosmological horizon radius ($\lds < r_{\text{eff}} < \infty$), this formula corresponds to the result of the $T\bar{T}$-deformed dS/CFT duality. When the effective radius of the sphere is less than the cosmological horizon radius ($0 < r_{\text{eff}} < \lds$), this formula agrees with the results of dS/dS holography and half-dS holography. Through an analysis of entanglement entropy compliance with strong subadditivity (SSA) and boosted strong subadditivity (BSSA) in both dS/dS holography and half-dS holography, we demonstrate that the dual field theory$-$residing on a timelike slice in dS spacetime$-$exhibits nonlocality in these holographic frameworks~\cite{Geng:2019ruz, Geng:2020kxh,Kawamoto:2023nki}.

	This paper is organized as follows: In Section~\ref{section2}, we review the derivation of trace flow equations in AdS/CFT with Euclidean signature and in dS/CFT with Lorentzian signature. Subsequently, we demonstrate the analytic continuation from the AdS/CFT result to the dS/CFT counterpart through double Wick rotation. In Section~\ref{The_spherical_partition_function_method_for_(A)dS}, we review the computation of entanglement entropy for $T\bar{T}$-deformed CFTs, presenting general expressions applicable to both AdS and dS holography. In Section~\ref{(A)dS/Poincaré case}, we calculate the non-perturbative results for the entanglement entropy of arbitrary single spatial intervals in $T\bar{T}$-deformed theories on general manifolds by employing the CHM map.  We further investigate the impact of $T\bar{T}$ deformations in finite temperature and finite size systems. In Section~\ref{(A)dS/EOW brane case}, we examine the non-locality of the dual field theory in dS/dS and half-dS holography by scrutinizing SSA and BSSA. Finally, we summarize our findings and discuss potential future directions in Section~\ref{Conclusion and Discussion}.

	\section{Holographic $T\bar{T}$ Flow}
	\label{section2}
	\noindent 
	In this section, we review the holographic derivation of the trace flow equation for the $T\bar{T}$ deformation~\cite{Shyam:2017znq,Hartman:2018tkw,Kraus:2018xrn}. We then show that the trace flow equation in the dS/CFT duality can be obtained from its AdS counterpart via a double Wick rotation.

	\subsection{$T\bar{T}$ Flow in AdS/CFT}
	\label{adsse}
	\noindent 
	The action for Einstein gravity in Euclidean AdS$_3$ (EAdS$_3$) spacetime $\mathcal{M}$, with a boundary $\partial\mathcal{M}$ at a finite radial cutoff, is given by
	\begin{equation}
		\label{action}
		S = \frac{1}{16\pi G}\int_{\mathcal{M}} d^3x \sqrt{g} \left(\mathcal{R}^{(3)} + \frac{2}{\lads^2}\right) + \frac{1}{8\pi G}\int_{\partial\mathcal{M}} d^2x \sqrt{\gamma} \left(K - \frac{1}{\lads}\right),
	\end{equation}
	where $K$ denotes the trace of the extrinsic curvature of the timelike boundary $\partial\mathcal{M}$. The quasi-local stress tensor~\cite{Balasubramanian:1999re} is defined as the variation of the on-shell action with respect to the induced boundary metric $\gamma^{ij}$,
	\begin{equation}
		\label{stresstensor}
		T_{ij} = \frac{2}{\sqrt{\gamma}} \frac{\delta S}{\delta \gamma^{ij}}
		= \frac{1}{8\pi G} \left(K_{ij} - K\gamma_{ij} + \frac{1}{\lads}\gamma_{ij}\right).
	\end{equation}
	The EAdS$_3$ metric in Fefferman-Graham~\cite{FeffermanGraham1985} coordinates is
	\begin{equation}
		\label{FGgauge}
		ds^2 = \frac{\lads^2}{r^2} \left(dr^2 + h_{ij} dx^i dx^j\right), \quad \gamma_{ij} = \frac{\lads^2}{r_c^2} h_{ij}.
	\end{equation}
	In these coordinates, the spacetime is foliated by hypersurfaces labeled by the radial coordinate $r$. The extrinsic curvature of the hypersurface $\Sigma_{r_c}$ at $r = r_c$ is given by
	\begin{align}
		K_{ij} = \frac{1}{2} n^r \partial_r \left.\left(\frac{\lads^2}{r^2} h_{ij}\right)\right|_{r=r_c} = \frac{r}{2\lads} \partial_r \left.\left(\frac{\lads^2}{r^2} h_{ij}\right)\right|_{r=r_c},
	\end{align}
	where $n^r$ is the outward-pointing normal vector to the hypersurface $\Sigma_{r_c}$. The $(r,r)$ component of Einstein's equations is
	\begin{equation}
		\label{GC}
		K^2 - K_{ij}K^{ij} = \mathcal{R}^{(2)} + \frac{2}{\lads^2},
	\end{equation}
	where $\mathcal{R}^{(2)}$ is the Ricci scalar of the hypersurface $\Sigma_{r_c}$. Substituting the expression for $K_{ij}$ from eq.~\eqref{stresstensor} into eq.~\eqref{GC} yields the trace flow equation
	\begin{equation}
		\label{traceflow}
		T^{i}_i = -\frac{\lads}{16\pi G}\mathcal{R}^{(2)} - 4\pi G\lads \left[T^{ij}T_{ij} - (T^{i}_i)^2\right].
	\end{equation}

	Equation~\eqref{GC} is equivalent to the Hamiltonian constraint $H=0$ in the canonical form of Einstein's gravity theory~\cite{Arnowitt:1962hi}, with the Hamiltonian given by
	\begin{equation}
		\label{hami}
		H = \frac{16\pi G}{\sqrt{\gamma}} \left[\Pi^{ij}\Pi_{ij} - ({\Pi^i}_i)^2\right] + \frac{\sqrt{\gamma}}{16\pi G}\left(\mathcal{R}^{(2)} + \frac{2}{\lads^2}\right) = 0,
	\end{equation}
	where $\Pi^{ij}$ is the canonical momentum defined as
	\begin{equation}
		\Pi^{ij} = \frac{\sqrt{\gamma}}{16\pi G}(K\gamma^{ij} - K^{ij}).
	\end{equation}
	In the canonical form of quantum gravity with Euclidean signature, the wavefunction of the induced metric $\gamma_{ij}$ on the hypersurface $\Sigma_{r_c}$ can be written as $\Psi \equiv \text{e}^{-S}$, where $S$ is the classical action of Einstein gravity with the Gibbons-Hawking-York boundary term. After renormalization by introducing a counterterm, the wavefunction takes the form~\cite{Witten:2022xxp}
	\begin{equation}
		\label{conicaltrs}
		\Psi = \exp\left(-\frac{1}{8\pi G\lads}\int_{\partial\mathcal{M}} d^2x \sqrt{\gamma}\right) \hat{\Psi}.
	\end{equation}
	This corresponds to performing a canonical transformation on the conjugate momentum
	\begin{equation}
		\Pi^{ij} \to \Pi^{ij} + \frac{\sqrt{\gamma}}{16\pi G \lads} \gamma^{ij}.
	\end{equation}
	The Wheeler-DeWitt equation $H\hat{\Psi} = 0$~\cite{DeWitt:1967yk} in terms of the original conjugate momentum becomes
	\begin{equation}
		\label{hamiexpand}
		\begin{aligned}
			H\hat{\Psi} &= \left(\frac{16\pi G}{\sqrt{\gamma}}(\Pi^{ij}\Pi_{ij} - ({\Pi^i}_i)^2) - \frac{2}{\lads}\Pi + \frac{\sqrt{\gamma}}{16\pi G}\mathcal{R}^{(2)}\right)\hat{\Psi}, \\
			&= \sqrt{\gamma}\left(4\pi G(T^{ij}T_{ij} - (T^{i}_i)^2) + \frac{1}{\lads}T^{i}_i + \frac{\mathcal{R}^{(2)}}{16\pi G}\right)\hat{\Psi} = 0,
		\end{aligned}
	\end{equation}
	where in the second line we have used the relation $T^{ij} = -\frac{2}{\sqrt{\gamma}}\Pi^{ij}$. This aligns with the trace flow equation \eqref{traceflow}. To match this with the CFT trace flow equation
	\begin{equation}
		\label{eq:traceflowequ-CFT}
		T^{i}_i = -\frac{c}{24\pi} \mathcal{R}^{(2)} - 4\pi \lambda T\bar{T},  
	\end{equation}
	we require the central charge $c$ and the deformation parameter $\lambda$ to be
	\begin{equation}
		\text{AdS}: \quad c = \frac{3\lads}{2G} = c_\ads, \quad \lambda = 8G\lads = \lambda_\ads.
	\end{equation}

	\subsection{$T\bar{T}$ Flow in dS/CFT}
	\noindent 
	In the dS/CFT duality, the holographic direction corresponds to time, so the dS metric in Fefferman-Graham coordinates is given by
	\begin{equation}
		\label{FGdsgauge}
		ds^2 = \frac{\lds^2}{r^2}(-dr^2 + h_{ij}dx^idx^j), \quad \gamma_{ij} = \frac{\lds^2}{r_c^2} h_{ij},
	\end{equation}
	where $r$ is the holographic coordinate. The holographic boundary $\Sigma_{r_c}$, determined by $r = r_c$, is spacelike. The bulk geometry has a Lorentzian signature, and its gravitational partition function is assumed to be the Bunch-Davies wavefunction~\cite{Hikida:2022ltr,Doi:2022iyj}. Thus, according to the holographic dictionary, the partition function of the CFT is
	\begin{equation}
		\Psi_{\text{dS}} = Z_{\text{QFT}} = e^{i S_{\rm dS}},
	\end{equation}
	where $S_{\rm dS}$ is the on-shell action of Einstein gravity with the dS geometry,
	\begin{equation}
		S_{\rm dS} = \frac{1}{16\pi G}\int_{\mathcal{M}} d^3x \sqrt{-g} \left(\mathcal{R}^{(3)} - \frac{2}{\lds^2}\right) + \frac{1}{8\pi G}\int_{\partial\mathcal{M}} d^2x \sqrt{\gamma} \left(K - \frac{1}{\lds}\right).
	\end{equation}
	The trace flow equation for the dS/CFT case can be derived using a method parallel to that of the AdS/CFT case. First, we define the renormalized wavefunction by introducing boundary counterterms, corresponding to a canonical transformation on the conjugate momentum
	\begin{equation}
		\label{conicaltrsds}
		\Psi_{\rm dS} = \exp\left(\frac{i}{8\pi G\lds}\int_{\partial\mathcal{M}} d^2x \sqrt{\gamma}\right) \hat{\Psi}_{\rm dS}, \quad 
		\Pi^{ij}_{\rm dS} \to \Pi^{ij}_{\rm dS} + \frac{\sqrt{\gamma}}{16\pi G \lds} \gamma^{ij}.
	\end{equation}
	The Wheeler-DeWitt equation for the renormalized wavefunction, expressed in terms of the original conjugate momentum, is
	\begin{equation}
		\label{hamiexpandds}
		\begin{aligned}
			H\hat{\Psi}_{\rm dS} &= \left[\frac{16\pi G}{\sqrt{\gamma}}(\Pi^{ij}\Pi_{ij} - ({\Pi^i}_i)^2) - \frac{2}{\lds}\Pi - \frac{\sqrt{\gamma}}{16\pi G}\mathcal{R}^{(2)}\right]\hat{\Psi}_{\rm dS} \\
			&= \sqrt{\gamma}\left(-4\pi G(T^{ij}T_{ij} - (T^{i}_i)^2) - \frac{i}{\lds}T^{i}_i - \frac{\mathcal{R}^{(2)}}{16\pi G}\right)\hat{\Psi}_{\rm dS} = 0,
		\end{aligned}
	\end{equation}
	where $T^{ij} = -\frac{2i}{\sqrt{\gamma}}\Pi^{ij}$. This leads to the $T\bar{T}$ trace flow equation for the dS/CFT being
	\begin{equation}
		\label{dstracec}
		T^{i}_i = \frac{i\lds}{16\pi G}\mathcal{R}^{(2)} + i4\pi G\lds\left(T^{ij}T_{ij} - (T^{i}_i)^2\right).
	\end{equation}
	By comparing this with the field theory equation, the correspondences between the Brown-Henneaux central charge, the deformation parameter, and the gravitational quantities for the dS spacetime are given by
	\begin{equation}
		\label{dsconstrant}
		c = -i\frac{3\lds}{2G} = -ic_\ds\,, \quad 
		\lambda = -i8G\lds = -i\lambda_\ds\,.
	\end{equation}

	\subsection{Double Wick Rotation to dS/CFT
		\label{analytical_continuation11}}
	\noindent 
	In this subsection, we demonstrate that the trace flow equation for the dS/CFT, given by eq.~\eqref{dstracec}, can be derived by performing a double Wick rotation on the trace flow equation for the AdS/CFT, given by eq.~\eqref{traceflow}. The double Wick rotation is defined by
	\begin{equation}
		\label{analytical_continuation}
		\lads \to -i\lds, \quad r \to -ir.
	\end{equation}
	The transformations of various components in the partition function under the double Wick rotation are summarized in Table~\ref{tab:exampleds}\footnote{Note that for the general dS/CFT, where the CFT resides on the infinite spacelike boundary, the external curvature must be defined using the inward normal vector.}.
	\begin{table}[htbp]
		\centering
		\small
		\caption{Analytical continuation from EAdS to dS}
		\label{tab:exampleds}
		\begin{tabular}{|c|c|c|}
			\hline
			EAdS & dS & Analytical Continuation \\
			\hline
			$n_a^{(\text{AdS})} = \frac{\lads}{r}(dr)_a$ & $n_a^{(\text{dS})} = \frac{\lds}{r}(dr)_a$ & $n_a^{(\text{AdS})} \to -i n_a^{(\text{dS})}$ \\
			\hline
			$n^a_{(\text{AdS})} = \frac{r}{\lads}\left(\frac{\partial}{\partial r}\right)^a$ & $n^a_{(\text{dS})} = -\frac{r}{\lds}\left(\frac{\partial}{\partial r}\right)^a$ & $n^a_{(\text{AdS})} \to -i n^a_{(\text{dS})}$ \\
			\hline
			$K_{ab}^{(\text{AdS})} = \frac{1}{2} n^r \partial_r g_{ab}$ & $K_{ab}^{(\text{dS})} = -\frac{1}{2} n^r \partial_r g_{ab}$ & $K_{ab}^{(\text{AdS})} \to i K_{ab}^{(\text{dS})}$ \\
			\hline
			$\epsilon_{abc}^{(\text{AdS})} = \sqrt{g}\,(dr)_a \wedge (d\tau)_b \wedge (dx)_c$ & $\epsilon_{abc}^{(\text{dS})} = \sqrt{-g}\,(dr)_a \wedge (d\tau)_b \wedge (dx)_c$ & $\epsilon_{abc}^{(\text{AdS})} \to -i \epsilon_{abc}^{(\text{dS})}$ \\
			\hline
			$\epsilon_{ab}^{(\text{AdS})} = n^a_{(\text{AdS})} \epsilon_{abc}^{(\text{AdS})}$ & $\epsilon_{ab}^{(\text{dS})} = n^a_{(\text{dS})} \epsilon_{abc}^{(\text{dS})}$ & $\epsilon_{ab}^{(\text{AdS})} \to -\epsilon_{ab}^{(\text{dS})}$ \\
			\hline
			$\mathcal{R}_{(\text{AdS})}^{(3)} = -\frac{6}{\lads^2}$ & $\mathcal{R}_{(\text{dS})}^{(3)} = \frac{6}{\lds^2}$ & $\mathcal{R}_{(\text{AdS})}^{(3)} \to \mathcal{R}_{(\text{dS})}^{(3)}$ \\
			\hline
			$S_{ct}^{(\text{AdS})}$ & $S_{ct}^{(\text{dS})}$ & $S_{ct}^{(\text{AdS})} \to -i S_{ct}^{(\text{dS})}$ \\
			\hline
		\end{tabular}
	\end{table}
	By applying the transformation rules in Table~\ref{tab:exampleds} to the on-shell action of AdS gravity~\eqref{action}, we obtain the on-shell action for dS gravity, multiplied by an imaginary factor $i$:
	\begin{align}
		S_\ads &\rightarrow -\frac{i}{16\pi G}\int_{\mathcal{M}} d^3x \sqrt{-g} \left(\mathcal{R}^{(3)} - \frac{2}{\lds^2}\right) - \frac{1}{8\pi G}\int_{\partial\mathcal{M}} d^2x \sqrt{\gamma} \left(iK - \frac{i}{\lds}\right) \nonumber \\
		& = -i S_\ds.
		\label{actiontransformation}
	\end{align}
	Under the saddle point approximation, the double Wick rotation establishes that the wavefunction of AdS gravity, given by $\Psi = e^{-S_{\text{AdS}}}$, corresponds to the Bunch-Davies wavefunction of dS gravity, expressed as $\Psi_{\text{dS}} = e^{i S_{\text{dS}}}$. As a result, eqs.~\eqref{conicaltrs} and \eqref{hamiexpand} correspond to eqs.~\eqref{conicaltrsds} and \eqref{hamiexpandds}, respectively. This means that the trace flow equation~\eqref{dstracec} in the dS/CFT can be directly derived by applying the double Wick rotation to the trace flow equation~\eqref{traceflow} in the AdS/CFT. At the same time, it should be noted that although we have linked the WDW constraints in dS spacetime with the WDW constraints in AdS spacetime through the double Wick rotation, the Hilbert space of dS quantum gravity may not be derived by the analytical continuation from that of AdS\footnote{Thanks to the referee for pointing out the limitations of the analytical continuation method here.}. This is because the radial wave functional in AdS spacetime obeys regularity at $r = 0$, fixing the asymptotic form at $r = \infty$ to the phase factor times the partition function of the CFT. This constraint does not apply in dS spacetime\cite{Chakraborty:2023yed,Cianfrani:2013oja,Chakraborty:2023los}.

	\section{The Replica Method for the $T\bar{T}$-Deformed Field Theory}
	\label{The_spherical_partition_function_method_for_(A)dS}
	\noindent 
	This section reviews the replica trick for calculating the entanglement entropy of the subregion between two antipodal points on a two-dimensional sphere~\cite{Donnelly:2018bef, Chen:2023eic}. The sphere serves as the background manifold for the $T\bar{T}$-deformed CFT, with its radius linked to the deformation parameter from a holographic perspective. In the dS/CFT correspondence, the CFT resides on a future spacelike boundary, which is also a two-dimensional sphere in global dS coordinates. Thus, the replica method applied in the AdS case is equally applicable in the dS case.

\begin{figure}[H]
		\centering
		\includegraphics[scale=0.8]{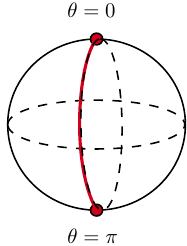}
		\caption{The meridian at $\phi = \text{constant}$ with endpoints at antipodal points defines the subregion $A$.}
		\label{figsphere}
	\end{figure}

	The metric that describes the boundary two-dimensional sphere in both AdS and dS cases is
	\begin{equation}
		\label{sphere}
		ds^2 = \gamma_{ab} dx^a dx^b = r^2 \left(d\theta^2 + \sin^2\theta \, d\phi^2\right),
	\end{equation}
	where $r$ represents the effective radius of the boundary sphere. The replica trick is used to compute the entanglement entropy of the subregion $A$, defined between the antipodal points $\theta = 0$ and $\theta = \pi$ at $\phi = \text{constant}$, as illustrated in figure~\ref{figsphere}.
	By performing $n$ copies along the meridian at $\phi = \text{constant}$, the $n$-replica geometry is obtained~\cite{Dong:2023bax, Lewkowycz:2013nqa} by
	\begin{equation}
		ds_n^2 = r^2 \left(d\theta^2 + n^2 \sin^2\theta \, d\phi^2\right).
		\label{eq:metric-n-sphere}
	\end{equation}
	The entanglement entropy of the subregion $A$ is computed by taking the limit $n \rightarrow 1$ of the conical entropy~\cite{Headrick:2010zt, Hung:2011nu}
	\begin{equation}
		\label{eq:EE-CE}
		S_A = \lim_{n\to 1} \left(1 - n \frac{\partial}{\partial n}\right) \ln Z_n,
	\end{equation}
	where $Z_n$ is the partition function of the field theory on the $n$-replica geometry. Based on the definition of the stress tensor, the following expressions hold
	\begin{align}
		\frac{d}{dr} \ln Z_n
		& = -\int_{\mathbb{S}_n^2} d^2x \frac{\delta\mathcal{L}}{\delta g_{ab}} \frac{d g_{ab}}{dr}
		= -\frac{1}{r} \int_{\mathbb{S}_n^2} d^2x \sqrt{g} T^{ab} g_{ab},
		\label{eq:dlnz/dr}
	\end{align}
	and
	\begin{align}
		\frac{\partial}{\partial n} \ln Z_n
		& = -\int_{\mathbb{S}_n^2} d^2x \frac{\delta\mathcal{L}}{\delta g_{\phi\phi}} \frac{\partial g_{\phi\phi}}{\partial n}
		= -\frac{1}{n} \int_{\mathbb{S}_n^2} d^2x \sqrt{g} T^{\phi\phi} g_{\phi\phi}.
		\label{eq:dlnzn/dn}
	\end{align}
	The trace flow equation and the conservation equation determine the stress tensor as follows:
	\begin{align}
		\label{eq:trace-flow}
			T^{a}_a = -\frac{c}{24\pi} R^{(2)} - \frac{\pi \lambda}{2} \left[T^{ab}T_{ab} - (T^{a}_a)^2\right], \\
			\nabla_a T^{a}_b = 0.
	\end{align}
	Due to the maximal symmetry of the geometry, the quasi-local stress tensor is assumed to take the form
	\begin{align}
		\label{eq:T-isotropy}
		T_{ab} = \alpha \gamma_{ab}, \quad
		T^{\theta}_\theta = T^{\phi}_\phi = \frac{1}{2}\text{tr}\, T.
	\end{align}
	Under these assumptions, it follows that $\frac{d\ln Z_n}{dr}\big|_{n=1} = -8\pi r \alpha$, and the entanglement entropy of the subregion $A$ is given by
	\begin{equation}
		S_A = \lim_{n\to 1} \left(1 - \frac{r}{2} \frac{d}{dr}\right) \ln Z_n.
	\end{equation}
	In this case, the parameter $\alpha$ determines the stress tensor, which is uniquely defined by the trace flow equations.

	\subsection{AdS spacetime}
\begin{figure}[H]
		\centering
		\subfigure[$C(r)$]{
			\includegraphics[width=0.4\textwidth]{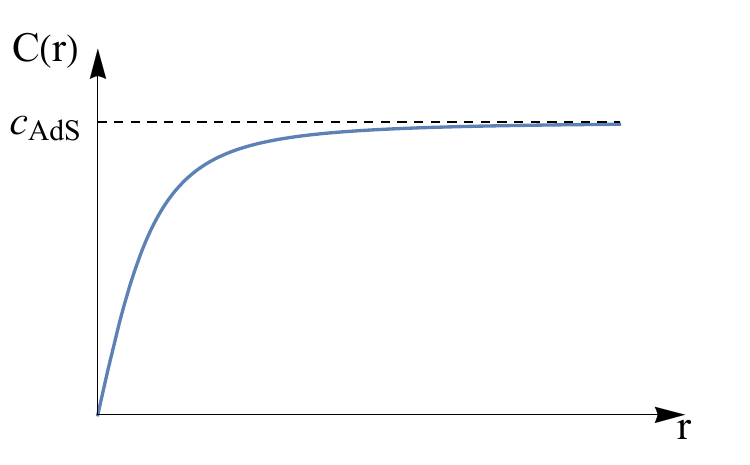}}
		\hspace{1in}
		\subfigure[$C(\lambda)$]{
			\includegraphics[width=0.4\textwidth]{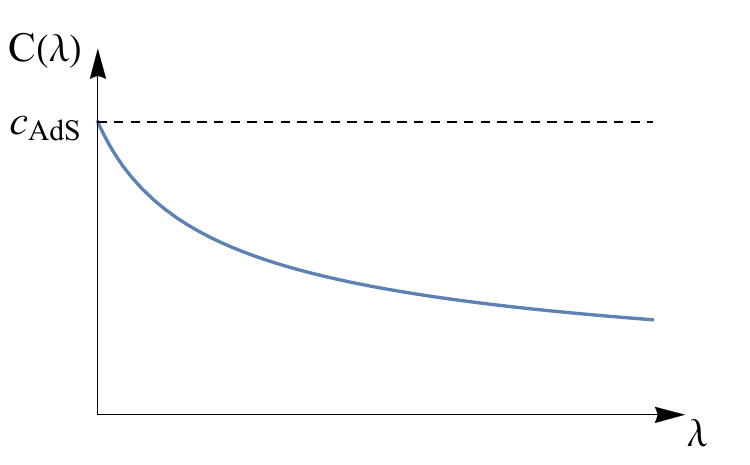}}
		\caption{The $C$-function of the AdS spacetime. We use $\lambda=\frac{\lambda_\ads}{r^2}$ to represent the deformation parameter. As the boundary cutoff $r = r_c$ is pushed further into the bulk, the deformation parameter increases and the theory deviates from the CFT increasingly.}
		\label{fig:The C-function of AdS spacetime.}
	\end{figure}

	\noindent
	By substituting the assumed form of the stress tensor $T_{ij} = \alpha \gamma_{ij}$ into eq.~\eqref{traceflow}, the equation for $\alpha$ is obtained as
	\begin{equation}
		8\pi G \lads \alpha^2 = 2\alpha + \frac{\lads}{16\pi G} \mathcal{R}^{(2)},
	\end{equation}
	where $\mathcal{R}^{(2)} = \frac{2}{r^2}$ for the two-dimensional sphere~\eqref{sphere}. The solution to this equation is given by
	\begin{equation}
		\alpha = \frac{1}{\pi \lambda_\ads} \left( 1 - \sqrt{1 + \frac{\lambda_\ads c_\ads}{12r^2}} \right).
	\end{equation}
	Substituting this expression for $\alpha$ into eq.~\eqref{eq:dlnz/dr}, the $n$-replica partition function $Z_n$ can be computed by integrating with respect to $\alpha$:
	\begin{equation}
		\ln Z_n = \frac{c_\ads}{3} \tanh^{-1}\left(\frac{1}{\sqrt{1 + \frac{c_\ads \lambda_\ads}{12r^2}}}\right) + \frac{4r^2}{\lambda_\ads} \left(-1 + \sqrt{1 + \frac{c_\ads \lambda_\ads}{12r^2}}\right).
	\end{equation}
	The entanglement entropy of the subregion $A$ is then obtained as
	\begin{equation}
		\label{expressads}
		S_A = \lim_{n \to 1} \left(1 - \frac{r}{2} \frac{d}{dr}\right) \log Z_n = \frac{c_\ads}{3} \tanh^{-1}\left(\frac{1}{\sqrt{1 + \frac{c_\ads \lambda_\ads}{12r^2}}}\right).
	\end{equation}
	The entanglement entropy between two antipodal points on a sphere reflects the number of degrees of freedom in the field theory defined on the sphere~\cite{Casini:2012ei}. Casini and Huerta, using the principles of unitarity and locality in quantum field theory (QFT), demonstrated that the $C$-function, which measures the degrees of freedom, exhibits a monotonically decreasing behavior along the renormalization group flow for unitary and Lorentz-invariant QFTs~\cite{Casini:2004bw, Barnes:2004jj, Gukov:2015qea}. This validates Zamolodchikov's $C$-theorem~\cite{Zamolodchikov:1986gt}. Since the $T\bar{T}$ trace flow equation is equivalent to the holographic renormalization group flow equation, the renormalization group flow can be controlled by adjusting the radius of the sphere~\cite{deBoer:1999tgo, Verlinde:1999xm, Heemskerk:2010hk,  Nishioka:2018khk,Grieninger:2023knz,  Kamei:2024iei}. In the AdS case, the $C$-function can be constructed through the entanglement entropy.
	\begin{equation}
		\label{cadsfff}
		C(r) \equiv 3r S_A^{\prime}(r) = c_\ads \frac{1}{\sqrt{1 + \frac{c_\ads \lambda_\ads}{12r^2}}}\,.
	\end{equation}
	As illustrated in figure~\ref{fig:The C-function of AdS spacetime.}, the $C$-function decreases monotonically as $r$ decreases. Additionally, as $r$ approaches infinity, the $C$-function converges to the central charge of the two-dimensional CFT, indicating that the renormalization group flow approaches the UV fixed point~\cite{Nakayama:2013is}. This behavior is in agreement with the $C$-theorem.

	\subsection{dS spacetime}
	\label{The dS case}
\begin{figure}[H]
			\centering
			\subfigure[dS/CFT]{
				\label{ds1}
				\includegraphics[scale=0.9]{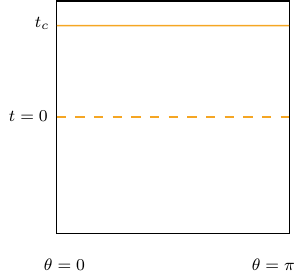}}
			\hspace{1in}
			\subfigure[dS/dS]{
				\label{ds2}
				\includegraphics[scale=0.9]{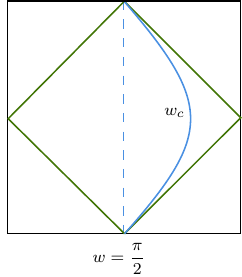}}
			\caption{The Penrose diagram for de Sitter spacetime. \ref{ds1}The orange solid line represents the finite cutoff at $t=t_c$, while the orange dashed line represents $t=0$. \ref{ds2}The green squares represent the boundaries at $w=0, \pi$ in dS/dS coordinates. The blue dashed line represents $w=\pi/2$, and the blue solid line represents the finite cutoff at $w=w_c$.}
			\label{dsstatic}
		\end{figure}

	\noindent
		In the context of dS spacetime, two primary holographic models are considered: dS/CFT and dS/dS holography, as illustrated in figures~\ref{ds1} and~\ref{ds2}. The dS/CFT duality is conventionally formulated in global dS coordinates, whose metric is expressed as
		\begin{equation}
			ds^2 = \lds^2 \left[-dt^2 + \cosh^2 t \left(d\theta^2 + \sin^2\theta \, d\phi^2\right)\right],
		\end{equation}
		where the $T\bar{T}$-deformed CFT resides on a boundary at a finite cutoff $t=t_c$ and the radius of the boundary sphere, $r = \lds \cosh t_c$, exceeds $\lds$ for any finite cutoff except $t_c=0$. For dS/dS holography, the metric in dS/dS coordinates is
		\begin{equation}
			ds^2 = \lds^2 \left[dw^2 + \sin^2 w \left(-dt^2 + \cosh^2 t \, d\phi^2\right)\right].
		\end{equation}
		In the Euclidean signature, the dual field theory is defined on a two-dimensional sphere with a radius $r = \lds \sin w_c$. Notably, for any finite radial cutoff $w_c$, the sphere's radius $r$ remains strictly less than $\lds$. The critical condition $r = \lds$ corresponds to the central slices of dS spacetime, as depicted by the orange and blue dashed lines in Figure~\ref{dsstatic}. According to the surface/state correspondence~\cite{Miyaji:2015yva}, since the central slice in dS spacetime is an extremal surface devoid of internal entanglement, the dual field theory on it should exhibit non-locality, i.e., we expect the $C$-function to diverge.
	
\begin{figure}[H]
		\centering
		\subfigure[$C(r)$]{
			\includegraphics[width=0.4\textwidth]{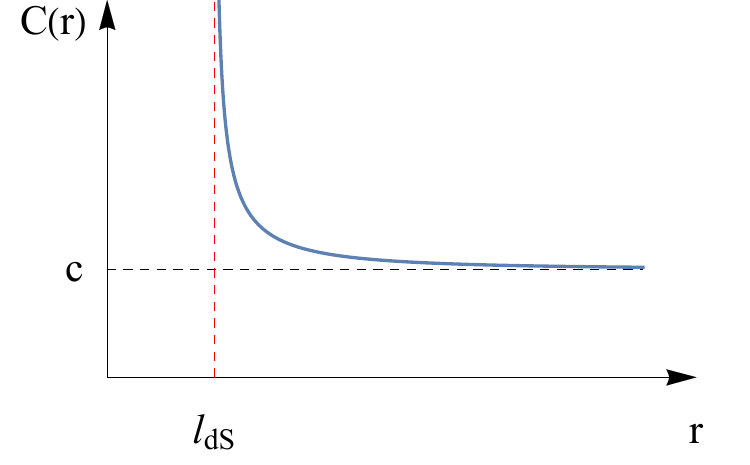}}
		\hspace{1in}
		\subfigure[$C(\lambda)$]{
			\includegraphics[width=0.4\textwidth]{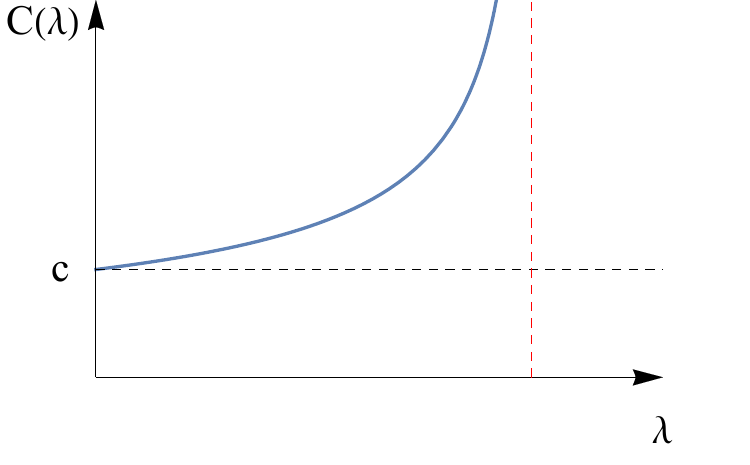}}
		\caption{The $C$-function of the dS spacetime at  $r>\lds$, and $\lambda=\frac{\lambda_\ds}{r^2}$. The red dashed line represents $r = \ell_{\text{dS}}$, that is $t_c=0$.
		}
		\label{fig:The C-function of dS spacetime.}
	\end{figure}

	\noindent\textbf{dS/CFT:}\\
	\indent For the dS/CFT duality, the assumed form of the quasi-local stress tensor~\eqref{eq:T-isotropy} is substituted into eq.~\eqref{dstracec} to determine the value of $\alpha$:
	\begin{equation}
		8\pi G \lds \alpha^2 = 2 i \alpha + \frac{\lds}{16\pi G} \mathcal{R}^{(2)}, \quad
		\alpha = \frac{i}{\pi \lambda_\ds} \left( 1 - \sqrt{1 - \frac{\lambda_\ds c_\ds}{12r^2}} \right).
	\end{equation}
	By integrating $\alpha$ with respect to $r$, the $n$-replica partition function is obtained as
	\begin{equation}
		\ln Z_n = -\frac{i c_\ds}{3} \tanh^{-1} \left( \frac{1}{\sqrt{1 - \frac{c_\ds \lambda_\ds}{12r^2}}} \right)
		+ \frac{i 4r^2}{\lambda_\ds} \left( -1 + \sqrt{1 - \frac{c_\ds \lambda_\ds}{12r^2}} \right).
	\end{equation}
	The entanglement entropy between of $A$ is then given by
	\begin{equation}
		\label{dsdsentropy}
		S_A = \lim_{n \to 1} \left( 1 - \frac{r}{2} \frac{d}{dr} \right) \ln Z_n
		= -\frac{i c_\ds}{3} \tanh^{-1} \left( \frac{1}{\sqrt{1 - \frac{c_\ds \lambda_\ds}{12r^2}}} \right).
	\end{equation}
	In this case, the argument of the inverse hyperbolic function is greater than one, making $\tanh^{-1}\left( \frac{1}{\sqrt{1 - \frac{c_\ds \lambda_\ds}{12r^2}}} \right)$ a complex value. The positive branch of the inverse hyperbolic function is chosen to match the real part of the entanglement entropy in the dS/CFT. The final expression for the entanglement entropy in the dS/CFT correspondence is
	\begin{equation}
		S_A = -\frac{i c_\ds}{3} \coth^{-1} \left( \frac{1}{\sqrt{1 - \frac{c_\ds \lambda_\ds}{12r^2}}} \right) + \frac{\pi c_\ds}{6},
	\end{equation}
where the real part agrees with the results in refs.~\cite{Hikida:2022ltr,Doi:2023zaf, Narayan:2022afv, Narayan:2023zen}. As discussed in ref.~\cite{Doi:2022iyj}, the entanglement entropy in the dS/CFT can be interpreted as the pseudo entropy, which is viewed as the origin of the non-unitarity of the boundary field theory in the dS/CFT. Additionally, we still want to use the $C$-function to measure the degree of freedom of the field theory. However, due to the non-unitarity of the dual field theory, the $C$-function is also a purely imaginary number.
	\begin{equation}
		\label{cdsfun}
		C(r)=-i c_\ds\frac{1}{\sqrt{1-\frac{c_\ds\lambda_\ds }{12r^2}}}\,.
	\end{equation}
	For the undeformed CFT, we have $C(r\to\infty)=-i c_\ds=c$. But as $r\to \lds $, the imaginary part of the $C$-function diverges, as illustrated in figure~\ref{fig:The C-function of dS spacetime.}, as anticipated.
	
\begin{figure}[H]
		\centering
		\subfigure[$C(r)$]{
			\includegraphics[width=0.4\textwidth]{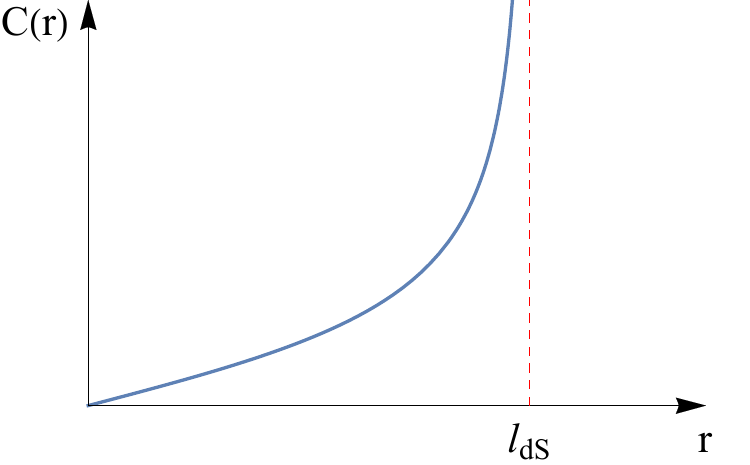}}
		\hspace{1in}
		\subfigure[$C(\lambda)$]{
			\includegraphics[width=0.4\textwidth]{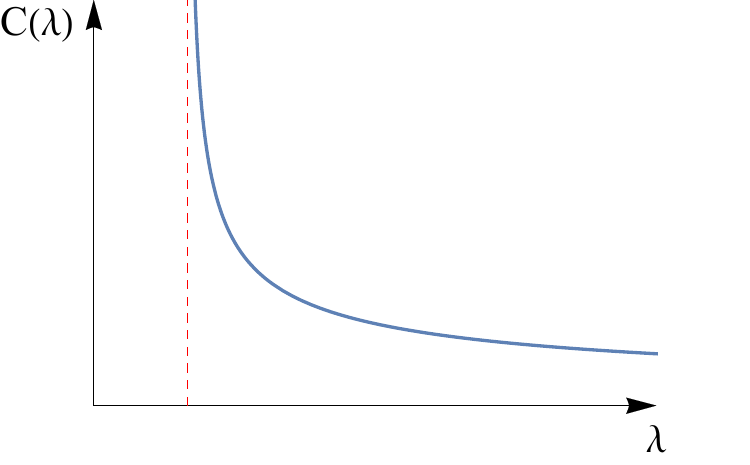}}
		\caption{The $C$-function of the dS spacetime at  $r<\lds$. The red dashed line indicates the ultraviolet boundary at $w_c=\frac{\pi}{2}$.}
		\label{fig:The C-function of rdS spacetime.}
	\end{figure}

\begin{figure}[H]
			\centering
			\includegraphics[scale=0.8]{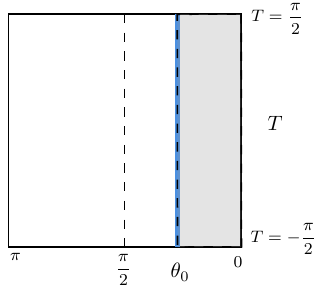}
			\caption{Half-dS holography. Here, we set $\cos T = \frac{1}{\cosh t}$. The dual field theory is defined on the timelike boundary $\theta = \theta_0$, and the bulk spacetime is represented by the shaded region.}
			\label{halfdsglobal}
		\end{figure}

	\noindent\textbf{dS/dS:}\\
	\indent In the dS/dS holography, the entanglement entropy and $C$-function have been analyzed in refs.~\cite{Gorbenko:2018oov}, with the results given as
	\begin{equation}
		\label{the ds timelike result}
		S_A = \frac{c_\ds}{3} \tan^{-1}\left(\frac{1}{\sqrt{\frac{c_\ds \lambda_\ds}{12r^2} - 1}}\right), \quad 
		C(r) = c_\ds \frac{1}{\sqrt{\frac{c_\ds \lambda_\ds}{12r^2} - 1}}.
	\end{equation}
	These two results can also be directly derived from expressions~\eqref{dsdsentropy} and~\eqref{cdsfun} within the dS/CFT duality by recognizing that the sphere's radius is smaller than the de Sitter radius $\lds$. As anticipated, the $C$-function diverges at the ultraviolet boundary $w_c=\frac{\pi}{2}$, as illustrated in figure~\ref{fig:The C-function of rdS spacetime.}.
		\noindent\textbf{Half-dS holography:}\\
		\indent Recently, a novel holographic framework, half-dS holography, has been proposed~\cite{Kawamoto:2023nki}. This model introduces a timelike boundary into global dS spacetime. The induced metric on this boundary is given by
		\begin{equation}
			ds^2 = \lds^2 (-dt^2 + \cosh^2 t \sin^2 \theta_0 d\phi^2).
		\end{equation}
		The dual field theory is defined on this timelike boundary, characterized by the condition $\theta = \theta_0$. To ensure that spacelike geodesics can probe deep into the bulk spacetime, the bulk region is restricted to $0 < \theta < \theta_0$, as indicated by the shaded area in figure~\ref{halfdsglobal}. Reference~\cite{Kawamoto:2023nki} proposes that the dual field theory in this model also exhibits non-locality. Specifically, when the timelike boundary is located at $\theta_0=\frac{\pi}{2}$, the subadditivity is consistently violated for $t\neq 0$, except for the subadditivity saturation at $t=0$. This position coincides with the ultraviolet boundary position $w_c=\frac{\pi}{2}$ in dS/dS holography, which corresponds to the black dashed line in figure~\ref{halfdsglobal} and the blue dashed line in figure~\ref{ds2}, respectively.
		The dual field theories in dS/dS holography and half-dS holography are both defined on timelike boundaries within dS spacetime and exhibit non-locality at the central slice. To this end, we will extend the replica method for calculating entanglement entropy between antipodal points on a sphere to compute the entanglement entropy between any spatial region and its complement on a general manifold in Section~\ref{(A)dS/Poincaré case}. Subsequently, we will apply the replica method to dS/dS holography and half-dS holography in Section~\ref{(A)dS/EOW brane case} to discuss the similarities and differences between these two holographic models.

	\section{Entanglement entropy for arbitrary single spatial intervals}
	\label{(A)dS/Poincaré case}
	\noindent 
	In section~\ref{The_spherical_partition_function_method_for_(A)dS}, the replica method is employed to calculate the entanglement entropy for a subsystem between two antipodal points on a sphere under $T\bar{T}$ deformation. This method applies only to subsystems along semicircular arcs between antipodal points. 
	While the entanglement entropy of a single interval subjected to $T\bar{T}$ deformation on a planar manifold can be determined through CHM mapping and endpoint integration of the entanglement region~\cite{Lewkowycz:2019xse,Deng:2023pjs}, a comprehensive calculation of the entanglement entropy for an arbitrary single interval under $T\bar{T}$ deformation on spherical and cylindrical manifolds remains absent\footnote{In ref.~\cite{Grieninger:2019zts}, the authors proposed an entanglement entropy expression for a general interval, formatted analogously to the entanglement entropy expression between antipodal points on a spherical manifold, by introducing an effective radius. However, a rigorous proof of this proposition is lacking.}. In this section, we will extend the replica method to general manifolds, including spherical and cylindrical manifolds, to compute the $T\bar{T}$ corrections to the entanglement entropy for any arbitrary single interval.

	\subsection{The generalization of the replica method}
	\label{subsec:CHM}	
\begin{figure}[H]
			\centering
			\includegraphics[scale=1.2]{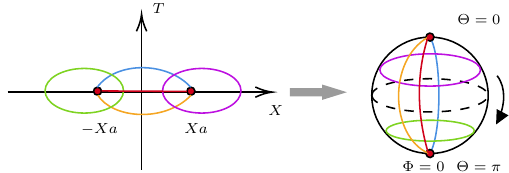}
			\caption{Mapping from the plane to the sphere. The north and south poles of the sphere correspond to the endpoints of the interval $A$ on the complex plane. The subregion $A$ corresponds to the meridian at $\Phi = 0$.}
			\label{mapads}
		\end{figure}

	\noindent In ref.~\cite{Lewkowycz:2019xse}, it is proposed that in the $T\bar{T}$-deformed theory, all dependence of the entanglement entropy on $\lambda$ originates from the contributions of the endpoints of the entanglement region. For the reader's convenience, we provide a concise review of the relevant derivation in Appendix~\ref{endpoints} and apply it to a planar manifold via the CHM map, yielding the expression for the entanglement entropy of a single interval in AdS Poincaré spacetime. Drawing inspiration from this, we can employ analogous steps in dS Poincaré spacetime to derive the entanglement entropy expression for a single spatial interval in the $T\bar{T}$-deformed CFT. For dS Poincaré spacetime, the metric is given by
		\begin{align}
			ds^2 = \frac{\lds^2}{\eta^2}(-d\eta^2+dX^2+dT^2)\,.
		\end{align}
		In this context, the $T\bar{T}$-deformed CFT resides on the two-dimensional complex plane $\eta=\eta_c$. The subregion $A$ is defined by $X \in [-X_a, X_a]$ at $T = 0$. It is mapped via the CHM transformation to a meridian $\Phi=0$ on a sphere with an effective radius $r_\eff=\frac{\lds X_a}{\eta_c}$, as illustrated in figure~\ref{mapads}. 
		\noindent The solution for the energy-momentum tensor on the $n$-replica sphere is given by 
		\begin{align}
			T^{\Theta}_\Theta
			&=\frac{i}{\pi\lambda_\ds}
			\left(1-\sqrt{1-\frac{\lambda_\ds c_\ds}{12r_\eff^2} - \frac{\lambda_\ds c_\ds}{12r_\eff^2}\left(\frac{1}{n^2}-1\right)\frac{1}{\sin^2\Theta}}\right),\nn\\
			T^{\Phi}_\Phi
			&=\frac{i}{\pi\lambda_\ds}
			\left(1-\frac{1-\frac{\lambda_\ds c_\ds}{12r_\eff^2}}{\sqrt{1-\frac{\lambda_\ds c_\ds}{12r_\eff^2}-\frac{\lambda_\ds c_\ds}{12r_\eff^2}\left(\frac{1}{n^2}-1\right)\frac{1}{\sin^2\Theta}}}\right)\,,
			\label{eq:T-nsphere-dS}
		\end{align}
		where the effective radius is $r_\eff=\frac{\lds X_a}{\eta_c}$. So we can obtain the $C$-function of the dS Poincaré spacetime
		\begin{align}
			C(r)=-\frac{ic_\ds}{3}\frac{1}{\sqrt{1-\frac{\eta_c^2 }{X_a^2}}}\,.
			\label{eq:EE-ds-poincare}
		\end{align}
		Finally, the entanglement entropy is given by
		\begin{align}
			S_A=-\frac{ic_\ds}{3}\coth^{-1}\left(\frac{1}{\sqrt{1-\frac{\eta_c^2 }{X_a^2}}}\right)
			+\frac{\pi c_\ds}{6}\simeq-\frac{ic_\ds}{3}\log\frac{2X_a}{\eta_c}+\frac{i c_\ds\eta_c^2}{12X_a^2}+\frac{\pi c_\ds}{6}\,.
			\label{eq:EE-poincare-ds}
		\end{align}
		In Appendix~\ref{appx:RT-dS/Poincaré}, we obtain consistent results by computing the area of the RT surface in the bulk spacetime with a finite cutoff $\eta_c$.

	To generalize the replica method for arbitrary holographic boundaries, we note that any two-dimensional manifold is locally conformally equivalent to the plane. Thus, its line element can be expressed as
	\begin{align}
		ds^2 = \omega(T, X)^2 (dT^2 + dX^2) \,.
		\label{eq:metric-conformal-plane}
	\end{align}
	Suppose the subregion $A$ in the coordinate system $(T, X)$ is a line segment with endpoints at $(T_a, -X_a)$ and $(T_a, X_a)$. To calculate the entanglement entropy of this subsystem, we map it onto the sphere using the transformation
	\begin{align}
		T &\rightarrow \frac{X_a \sin \Theta \sin \Phi}{1 + \sin \Theta \cos \Phi} + T_a \,, \quad
		X \rightarrow \frac{X_a \cos \Theta}{1 + \sin \Theta \cos \Phi} \,.
	\end{align}
	This mapping sends the endpoints $(T_a, -X_a)$ and $(T_a, X_a)$ on the plane to the antipodal points on the sphere. The line element~\eqref{eq:metric-conformal-plane} in these new coordinates becomes
	\begin{align}
		ds^2 = \frac{\omega(T(\Theta, \Phi), X(\Theta, \Phi))^2 X_a^2}{(1 + \cos \Phi \sin \Theta)^2} \left(d\Theta^2 + \sin^2 \Theta d\Phi^2 \right) \,,
		\label{equ:metric-comformal-CHM}
	\end{align}
	where $\Theta(T_a, X_a) = 0$ and $\Theta(T_a, -X_a) = \pi$. We then perform a conformal transformation to eliminate the conformal factor $\frac{\omega(T(\Theta, \Phi), X(\Theta, \Phi))}{1 + \cos \Phi \sin \Theta}$, resulting in a sphere with an effective radius $r_\eff$. If $\omega(T(\Theta,\Phi), X(\Theta,\Phi))$ is constant at both endpoints, this factor only contributes to the effective radius. At the endpoints $\Theta=0$ and $\pi$, the factor $1+\cos\Phi\sin\Theta$ is 1. Thus, near these endpoints, the metric~\eqref{equ:metric-comformal-CHM} approximates that of a sphere, with an effective radius of $\omega(T_a,X_a)X_a$. Once again, assuming that the integrand in \eqref{eq:d-Cfunction} is non-zero only in small neighborhoods near the endpoints, the conformal transformation is the identity map in these neighborhoods. Thus, we can directly use the partition function on the sphere with radius $r_\eff$ as the partition function on the manifold $\mathcal{M}$ within the subregion $A$. Within these small neighborhoods, we can use the solution for the energy-momentum tensor on the $n$-sphere to compute the integral, with the effective radius in the solutions~\eqref{eq:T-nphere-AdS} or \eqref{eq:T-nsphere-dS} replaced by $r_\mathrm{eff} = \omega(T_a,X_a)X_a$.
	
If \( \omega(T, X) \) takes distinct constant values at the two endpoints, denoted \( \omega_a \) and \( \omega_{-a} \), respectively, the contributions to the integral \eqref{eq:d-Cfunction} from these endpoints correspond to spheres with radii \( \omega_a X_a \) and \( \omega_{-a} X_a \). These integrals are therefore replaced by the partition function on spheres with radii \( \omega_a X_a \) and \( \omega_{-a} X_a \), respectively.


	\subsection{System with finite size}
	\label{(A)dS global coordinate}
	\noindent
	In the preceding subsection, we reviewed the CHM mapping method for computing the entanglement entropy of $T\bar{T}$-deformed CFTs within spatial intervals on a two-dimensional plane. Specifically, we calculated the entanglement entropy of a subsystem in dS Poincaré spacetime. We now extend this methodology to two-dimensional cylindrical and spherical manifolds to investigate whether the $T\bar{T}$ deformation induces corrections to the entanglement entropy in finite-size systems.

	\subsubsection{AdS case}
	\label{sec:EE-globalads}
\begin{figure}[H]
		\centering
		\includegraphics[scale=0.8]{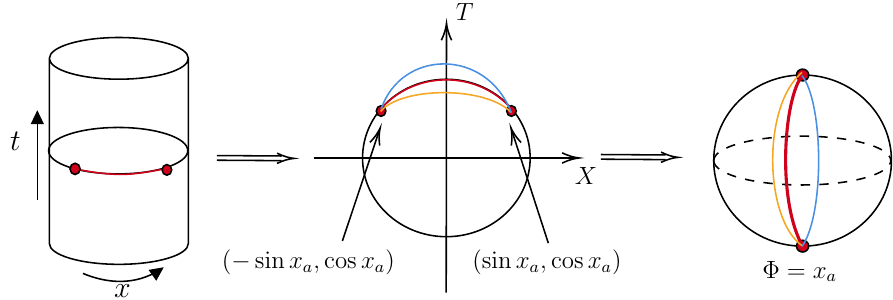}
		\caption{The CHM transformation for the AdS case. After applying two mappings, an arc on the cylinder with endpoints $(-x_a, x_a)$ is mapped to a meridian on the sphere at $\Phi = x_a$. The meridian is then replicated $n$ times by cutting it along the $\Phi$ direction. Using the replica trick, the entanglement entropy of the subregion $A$ is calculated.}
		\label{AdS global}
	\end{figure}

	\noindent
	The metric of EAdS$_3$ in global coordinates is
	\begin{align}
		ds_{\mathrm{EAdS}_3}^2 = \lads^2\left(d\rho^2 + \cosh^2\rho\, dt^2 + \sinh^2\rho\, dx^2\right)\,,
	\end{align}
	where $x \sim x + 2\pi$. At the boundary $\rho=\rho_c$, we can get the induced metric as 
	\begin{align}
		ds_{\mathrm{cyl}}^2 = \lads^2\sinh^2\rho_c\left(d\tilde{t}^2+dx^2\right)\,,
	\end{align}
	where $\tilde{t}=\frac{\cosh\rho_c}{\sinh\rho_c}t$. To compute the entanglement entropy of a subsystem in the subregion $A$ defined by the interval $(-x_a, x_a)$ on the time slice $t = 0$, we first map the cylinder onto the complex plane using the following transformations
	\begin{align}
		\tilde{t} \rightarrow \frac{1}{2} \ln (T^2 + X^2)\,, \quad
		x \rightarrow \frac{1}{2i} \ln \left( \frac{T + iX}{T - iX} \right)\,.
		\label{eq:cyl-to-plane}
	\end{align}
	Under this transformation, the endpoints $(0,-x_a)$ and $(0,x_a)$ of $A$ are mapped to $(T_a,-X_a)=(\cos x_a,-\sin x_a)$ and $(T_a,X_a)=(\cos x_a,\sin x_a)$, respectively. The line element in this coordinate becomes
	\begin{align}
		ds^2_{\mathrm{pl}} = \ell_\mathrm{AdS}^2 \sinh^2 \rho_c \frac{dT^2 + dX^2}{T^2 + X^2}\,.
	\end{align}
	Next, we apply the CHM transformation to map the plane onto a two-dimensional sphere, with the coordinate transformation
	\begin{align}
		T\rightarrow\frac{\sin x_a\sin\Theta \sin \Phi}{1+\sin\Theta\cos\Phi}+\cos x_a\,,\quad
		X\rightarrow\frac{\sin x_a\cos\Theta}{1+\sin\Theta\cos\Phi}\,.
	\end{align}
	This transformation maps the endpoints of the interval on the plane to antipodal points on the sphere. In these new coordinates, the metric becomes
	\begin{align}
		ds^2_{\Omega\mathbb{S}^2}
		&=\frac{\lads^2\sinh^2\rho_c\sin^2x_a}{(1+\cos(2x_a-\Phi)\sin\Theta)(1+\cos\Phi\sin\Theta)}\left(d\Theta^2+\sin^2\Theta d\Phi^2\right)\nn\\
		&=\Omega^2(\Theta,\Phi)r_\eff^2\left(d\Theta^2+\sin^2\Theta d\Phi^2\right)\,,
	\end{align}
	where the effective radius is $r_{\mathrm{eff}} = \ell_\mathrm{AdS} \sinh \rho_c \sin x_a$, and the conformal factor is $\Omega^{-2}(\Theta, \Phi) = (1 + \cos (2x_a - \Phi) \sin \Theta)(1 + \cos \Phi \sin \Theta)$. The mapping from the cylinder to the sphere is illustrated in figure~\ref{AdS global}.
	\noindent At the antipodal points $\Theta = 0$, $\pi$, the conformal factor simplifies to $\Omega = 1$, allowing us to directly compute the entanglement entropy. The resulting entanglement entropy for the subregion $A$ is given by
	\begin{align}
		S_A=&\frac{c_\ads}{3} \tanh^{-1}\left(\frac{1}{\sqrt{1 + \frac{c_\ads \lambda_\ads}{12r_\eff^2}}}\right)\nn\\
		\simeq&\frac{c_\ads}{3} \log (2 \sinh \rho_c \sin x_a) + \frac{c_\ads}{12 \sinh^2 \rho_c \sin^2 x_a}\,,
	\end{align}
	where the first term of the last line represents the entanglement entropy of the undeformed field theory for the interval $(-x_a,x_a)$ on the cylinder, with the UV cutoff $\epsilon = \frac{1}{2\sinh\rho_c}$. The second term accounts for the contribution from the $T\bar{T}$ deformation. This expression aligns with the perturbative results found in ref.~\cite{Chen:2018eqk}.

	\subsubsection{dS case}
\begin{figure}[H]
		\centering
		\includegraphics[scale=0.7]{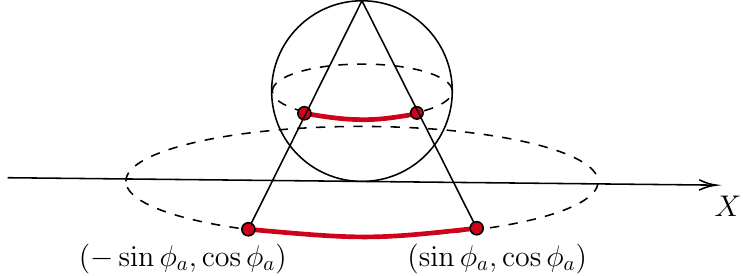}
		\caption{The CHM transformation of the global dS case.}
		\label{dS global}
	\end{figure}

	\noindent			
	For the global coordinates of the dS spacetime, the metric is given by
	\begin{align}
		\label{eq:metric-ds-global}
		ds^2 = \lds^2(-dt^2 + \cosh^2 t \, (d\theta^2 + \sin^2 \theta \, d\phi^2))\,.
	\end{align}
	After the $T\bar{T}$ deformation, its holographic boundary is a two-dimensional sphere determined by $t = t_c$. The subregion $A$ is defined by the arc $\theta = \pi/2$, $\phi \in (-\phi_a, \phi_a)$ on this two-dimensional sphere. As done previously, we begin by mapping the sphere to the complex plane using the transformation
	\begin{align}
		\theta = 2 \tan^{-1} \sqrt{T^2 + X^2}\,, \quad
		\phi = \frac{1}{2i} \ln \frac{T + iX}{T - iX}\,.
	\end{align}
	In this coordinate system, the endpoints of $A$ are $(\cos \phi_a, -\sin \phi_a)$ and $(\cos \phi_a, \sin \phi_a)$, and the line element becomes
	\begin{align}
		ds^2 = \frac{4 \lds^2 \cosh^2 t_c}{1 + X^2 + T^2} (dT^2 + dX^2)\,.
	\end{align}		
	As shown in figure~\ref{dS global}, after the above mapping, we map an arc on the sphere to an arc on the plane. We then apply the CHM transformation
	\begin{align}
		\label{cytoflat}
		T \rightarrow \frac{\sin \phi_a \sin \Theta \sin \Phi}{1 + \sin \Theta \cos \Phi} + \cos \phi_a\,,\quad
		X \rightarrow \frac{\sin \phi_a \cos \Theta}{1 + \sin \Theta \cos \Phi}\,,
	\end{align}
	which maps this subregion to the meridian $\Phi = \phi_a$ on the sphere. 
	\noindent
	The transformed line element is
	\begin{align}
		ds^2
		&= \frac{\lds^2 \cosh^2 t_c \sin^2 \phi_a}
		{(1 + \cos (\Phi - \phi_a) \cos\phi_a \sin\Theta)^2} (d\Theta^2 + \sin^2 \Theta d\Phi^2)\nn\\
		&= \Omega(\Theta, \Phi)^2 r_\eff^2 (d\Theta^2 + \sin^2 \Theta d\Phi^2)\,,
	\end{align}
	where $r_\eff = \lds \cosh t_c \sin \phi_a$ is the effective radius of the sphere, and the conformal factor is
	\begin{align}
		\Omega(\Theta, \Phi) = \frac{1}{1 + \cos (\Phi - \phi_a) \cos \phi_a \sin \Theta}\,.
	\end{align}
	At the antipodal points $\Theta = 0, \pi$, the conformal factor $\Omega(\Theta, \Phi) = 1$, allowing us to directly use the results in eq.~\eqref{eq:T-nsphere-dS} to calculate the entanglement entropy. The final result is
	\begin{align}
		\label{eq:EE-ds-global}
		S_A &=-\frac{ic_\ds}{3}\coth^{-1}\left(\frac{1}{\sqrt{1-\frac{c_\ds\lambda_\ds }{12r_\eff^2}}}\right)
		+\frac{\pi c_\ds}{6}\nn\\
		&\simeq -\frac{i c_\ds}{3} \log (2 \cosh t_c \sin \phi_a) + \frac{i c_\ds}{12 \cosh^2 t_c \sin^2 \phi_a} + \frac{\pi c_\ds}{6}.
	\end{align}
	This result is consistent with the area of the RT surface calculated in Appendix~\ref{appx:RT-globalds}, and also matches the findings of ref.~\cite{Chen:2023eic} when $\phi_a = \frac{\pi}{2}$.

	\subsection{System with finite temperature}
	\label{BTZ black hole and dS static space time}
	\noindent			
	In ref.~\cite{Chen:2018eqk}, the authors mentioned that entanglement entropy receives a first-order correction from the $T\bar{T}$ deformation only for the finite temperature case. Thus, in this subsection, we focus on investigating the entanglement entropy of the planar Ba$\tilde{\text{n}}$ados-Teitelboim-Zanelli (BTZ)~\cite{Banados:1992wn} black hole. For the Euclidean BTZ black hole, the bulk metric is given by
	\begin{align}
		ds^2 = \frac{R^2 - R_0^2}{\ell_\mathrm{AdS}^2} dt^2 + \frac{\ell_\mathrm{AdS}^2}{R^2 - R_0^2} dR^2 + R^2 dx^2\,,
		\label{eq:metric-BTZ}
	\end{align}
	where $x$ is the coordinate of the non-compact direction, and $t$ is periodic with $t \sim t + \beta$, where $\beta = \frac{2\pi \ell_\mathrm{AdS}^2}{R_0}$ is the inverse Hawking temperature. The induced metric on the holographic boundary $R = R_c$ is
	\begin{align}
		ds^2 = \frac{R_c^2 - R_0^2}{\ell_\mathrm{AdS}^2} \left( dt^2 + \frac{R_c^2 \ell_\mathrm{AdS}^2}{R_c^2 - R_0^2} dx^2 \right)
		\coloneqq \frac{R_c^2 - R_0^2}{\ell_\mathrm{AdS}^2} \left( dt^2 + d\tilde{x}^2 \right)\,,
	\end{align}
	where $\tilde{x} = \frac{R_c \ell_\mathrm{AdS}}{\sqrt{R_c^2 - R_0^2}} x$. We aim to compute the entanglement entropy of the subsystem within the interval $x \in (-x_a, x_a)$ at $t = 0$. The length of the interval is $L = \frac{2x_a R_c \lads}{\sqrt{R_c^2 - R_0^2}}$. To simplify the calculation, we redefine the coordinates by
	\begin{equation}
		\label{BTZtoads}
		R=R_0\cosh\rho,\quad t=\frac{\lads^2}{R_0}\phi=\frac{\beta}{2\pi}\phi,\quad x=\frac{\lads}{R_0}\tau
	\end{equation}
	to transform the metric~\eqref{eq:metric-BTZ} to the global coordinate of EAdS$_3$
	\begin{equation}
		ds^2 = \lads^2(\sinh^2\rho\, d\phi^2 + d\rho^2 + \cosh^2\rho\, d\tau^2).
	\end{equation}
	At this point, the chosen subregion corresponds to the interval $\tau \in (-\tau_a, \tau_a)$ at $\phi = 0$, where $\tau_a = \frac{R_0}{\lads} x_a$. The holographic boundary $R = R_c$ corresponds to $\rho = \rho_c$ with $\cosh\rho_c = \frac{R_c}{R_0}$. The induced metric on this boundary is 
	\begin{equation}
		ds^2 = \lads^2 \cosh^2 \rho_c (d\tau^2 + d\phi^2).
	\end{equation}
	Up to this point, the setup remains similar to that in sec~\ref{sec:EE-globalads}. According to eq.~\eqref{eq:cyl-to-plane}, the endpoints of the interval in the complex plane are $(T, X) = (e^{-\tau_a}, 0)$ and $(e^{\tau_a}, 0)$, and the line element in this coordinate becomes
	\begin{align}
		ds^2_{\mathrm{pl}} = \lads^2 \cosh^2 \rho_c \frac{dT^2 + dX^2}{T^2 + X^2}\,.
	\end{align}
	For the interval $X = 0$, with $T_a \in (e^{-\tau_a}, e^{\tau_a})$, the CHM map takes the form
	\begin{align}
		X \rightarrow \frac{\sinh \tau_a \sin\Theta \sin \Phi}{1+\sin\Theta\cos\Phi}\,,\quad
		T \rightarrow \frac{\sinh \tau_a \cos\Theta}{1+\sin\Theta\cos\Phi} + \cosh \tau_a\,.
	\end{align}
	
	\begin{figure}[H]
		\centering
		\includegraphics[scale=0.8]{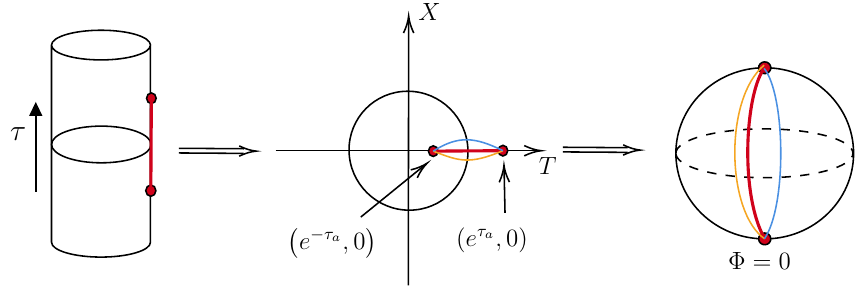}
		\caption{The CHM transformation of the BTZ black hole. In this case, the effective radius at the two endpoints is different.}
		\label{btz global}
	\end{figure}
	\noindent
	As depicted in figure~\ref{btz global}, under the CHM transformation, the interval along the cylindrical axis is mapped to the meridian at $\Phi = 0$ on the sphere. The resulting metric in these coordinates is given by
	\begin{equation}
		ds^2 = \frac{\lads^2 \cosh^2 \rho_c \sinh^2 \tau_a (d\Theta^2 + \sin^2 \Theta d\Phi^2)}{(1 + \cos\Phi\sin\Theta)(\cosh 2\tau_a + \cos\Phi\sin\Theta + \cos\Theta\sinh 2\tau_a)}.
		\label{eq:metric-inducedbtz-chm}
	\end{equation}
	
	In contrast to the previous situation, the effective radius $r_\eff$ at the endpoints $\Theta = 0$ and $\Theta = \pi$ is not equal. We can regard the constant factor $\lads \cosh \rho_c\sinh \tau_a$ in the metric~\eqref{eq:metric-inducedbtz-chm} as the role of radius ``$r$'' in eq.~\eqref{eq:dlnzn/dn-conformalsphere}, with the remaining angle-dependent part serving as the conformal factor $\Omega(\Theta, \Phi)$. Accordingly, the effective radii corresponding to the two endpoints are
	\begin{align}
		r_{\eff, \Theta = 0} &= r \Omega(0, \Phi) = \lads \cosh \rho_c \sinh \tau_a e^{-2\tau_a} = \lads \frac{R_c}{R_0} e^{\frac{-2\pi L \sqrt{1 - \left(\frac{R_0}{R_c}\right)^2}}{\beta}} \sinh \frac{\pi L \sqrt{1 - \left(\frac{R_0}{R_c}\right)^2}}{\beta}\,, \nn\\
		r_{\eff, \Theta = \pi} &= r \Omega(\pi, \Phi) = \lads \cosh \rho_c \sinh \tau_a e^{2\tau_a} = \lads \frac{R_c}{R_0} e^{\frac{2\pi L \sqrt{1 - \left(\frac{R_0}{R_c}\right)^2}}{\beta}} \sinh \frac{\pi L \sqrt{1 - \left(\frac{R_0}{R_c}\right)^2}}{\beta}\,.
	\end{align}
	As discussed at the end of Section~\ref{subsec:CHM}, we assume that the CHM mapping method is still applicable. However, the energy-momentum tensor in the integrals near different endpoints corresponds to solutions on the sphere with different radii. According to eq.~\eqref{eq:dC-conformalsphere}, we have
	\begin{align}
		\label{btzintergrate}
		r\frac{d}{dr}S_A(r)
		&=-\lim_{n\rightarrow1}2\pi\left[ r^2\Omega^2(0,\Phi)\int_0^{\Theta_0}\sin\Theta d\Theta\partial_n\tr T
		+r^2\Omega^2(\pi,\Phi)\int_{\pi-\Theta_0}^\pi\sin\Theta d\Theta\partial_n\tr T\right]\nn\\
		&=\frac{c}{6}\left(
		\frac{1}{\sqrt{1+\frac{c_\ads\lambda_\ads }{12 r^2\Omega^2(0,\Phi)}}}
		+\frac{1}{\sqrt{1+\frac{c_\ads\lambda_\ads}{12 r^2\Omega^2(\pi,\Phi)}}}\right)\,.
	\end{align}
	Integrating the above function with respect to $r$, the entanglement entropy is given by
	\begin{align}
		S_A = \frac{c_\ads}{6} \left[ \tanh^{-1} \left( \frac{1}{\sqrt{1 + \frac{c_\ads \lambda_\ads}{12 r_{\eff, \Theta = 0}^2}}} \right) 
		+ \tanh^{-1} \left( \frac{1}{\sqrt{1 + \frac{c_\ads \lambda_\ads}{12 r_{\eff, \Theta = \pi}^2}}} \right) \right]\,.
	\end{align}
	Substituting the radii $r_{\eff, \Theta}$ into this expression and expanding the result to second order in $R_0 / R_c$, 
	\begin{align}
		S_A 
		&\simeq \frac{c_\ads}{3} \left[ \ln \left( \frac{\beta R_c \sinh \frac{\pi L}{\beta}}{\pi \lads^2} \right) 
		- \frac{2\pi^3 \lads^4 L \coth \left( \frac{\pi L}{\beta} \right)}{R_c^2 \beta^3} 
		+ \frac{\pi^2 \lads^4 \left( 1 + \coth \left( \frac{2\pi L}{\beta} \right) \right)}{R_c^2 \beta^2} \right]\,,
	\end{align}
	where we have replaced $R_0$ with $\frac{2\pi \lads^2}{\beta}$. In the high-temperature limit $\beta \ll 1$, the last term is negligible compared to the second term\footnote{This result is consistent with the findings in refs.~\cite{Chen:2018eqk,Apolo:2023ckr} by identifying $R_c^2 = \frac{6 \lads^2}{\mu \pi c_\ads}$.}.


	\section{dS/dS holography and half-dS holography}
	\label{(A)dS/EOW brane case}
	\noindent
	In this section, we apply the replica method to dS/dS holography and half-dS holography. In contrast to the previously discussed examples, here, all boundary-induced metrics are dS metrics, and the dual field theory resides on dS spacetime. Previous studies on dS/dS holography and half-dS holography~\cite{Kawamoto:2023nki, Geng:2019ruz} suggest that the dual field theory exhibits non-locality. In this section, we will provide a more detailed discussion of entanglement entropy and non-locality, building upon these prior investigations.
		
\begin{figure}[H]
			\centering
			\includegraphics[scale=0.7]{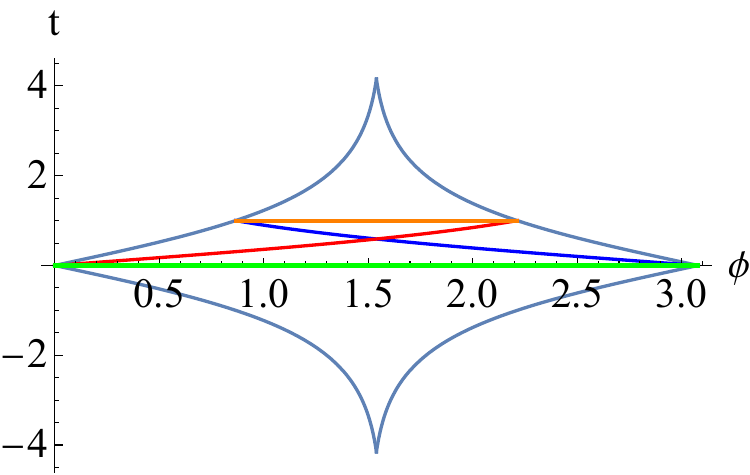}
			\caption{The causal structure of dS spacetime. The interval size we choose is $\Delta\phi=\frac{98\pi}{100}$.}
			\label{dsmin}
		\end{figure}

	Typically, the locality of a field theory is examined through strong subadditivity (SSA) and boosted strong subadditivity (BSSA). In Appendix~\ref{Strong subadditivity and boosted strong subadditivity}, we provide a concise review of the derivation of SSA and BSSA in flat spacetime. However, since the dual field theory in dS/dS holography and half-dS holography resides on dS space, we need to account for the effects of curved spacetime. The causal structure of dS spacetime exhibits a distinct geometry compared to the Minkowski causal diamond, owing to the metric's inherent time evolution, as illustrated in figure~\ref{dsmin}.
		In dS spacetime, since the SSA does not consider boost transformations, the intervals $A$ and $B$ remain on the $t=0$ slice. Consequently, the SSA is unaffected by the curvature of spacetime. Thus, the SSA in dS space is
		\begin{equation}
			S^{\prime\prime}(\Delta\phi) \leq 0.
		\end{equation}
		When deriving the BSSA, we must consider a small boost transformation of intervals $A$ and $B$. In figure~\ref{dsmin}, we represent interval $A$ with red segments, interval $B$ with blue segments, $A \cap B$ with orange segments, and $A \cup B$ with green segments. Considering that the entanglement entropy is a function of the interval size, we use the geodesic distance between two points as the interval length. We set the length of interval $A \cup B$ as $L_{A \cup B}=\Delta\phi$, and consider a small boost transformation that shrinks $\delta \phi$ in the $\phi$ direction. Then, the length of interval $A \cap B$ is
		\begin{equation}
			L_{A \cap B}=\cos^{-1}\left(1-2\sin^2\left(\frac{\Delta\phi}{2}-\delta\phi\right)\cosh\left(2\tanh^{-1}\left(\tan\frac{\delta\phi}{2}\right)\right)\right)\,.
		\end{equation}
		The lengths of intervals $A$ and $B$ are
		\begin{equation}
			L_A=L_B=\cos^{-1}\left( \cos(\Delta\phi-\delta\phi)\cosh\left(2\tanh^{-1}\left(\tan\frac{\delta\phi}{2}\right)\right)\right)\,.
		\end{equation}
		Taking the limit $\delta\phi\to 0$ and expanding formula~\eqref{bssa} in a series, we obtain the BSSA in dS space:
		\begin{equation}
			\label{dsdsBSSA}
			S^{\prime\prime}(\Delta\phi)+\frac{1}{\sin\Delta\phi}S^{\prime}(\Delta\phi)\leq 0\,.
		\end{equation}
		It is important to note that the preceding discussion is applicable only to the constant time slice at $t=0$. For cases where $t\neq 0$, we will directly use the entanglement entropy results obtained in the text to verify whether the SSA and BSSA are satisfied.

	\subsection{The dS/dS holography}
	\label{The result of dS/dS}
	\noindent
	For the dS/dS correspondence, the bulk metric is given by
	\begin{equation}
		\label{dsdsbulkmetric}
		ds^2=\lds ^2(dw^2+\sin^2w(-dt^2+\cosh^2t d\phi^2))\,,
	\end{equation}
	with the holographic boundary specified at $w=w_c$. The subregion $A$ is choosen as $t=t_0$, $\phi\in(-\phi_a,\phi_a)$. The boundary induced metric is
	\begin{equation}
		\label{dsdsinducedmetric}
		ds^2=\lds ^2\sin^2w_c(-dt^2+\cosh^2t d\phi^2)\,.
	\end{equation}
	In the Euclidean signature, after rescaling the time coordinate, the boundary metric can be written as
	\begin{equation}
		ds^2=\lds^2\frac{\sin^2w_c}{\cosh^2\tilde{t}}\left(d\tilde{t}^2+d\phi^2\right)\,,
	\end{equation}
	where $\tilde{t}$ is related to $t$ by the relation $\tilde{t}=\ln\frac{\cot\frac{t}{2}+1}{\cot\frac{t}{2}-1}$. This metric is conformally equivalent to the cylinder. As we have done in the last section, we can map it to the complex plane through the following transformations
	\begin{align}
		\tilde{t} \rightarrow \frac{1}{2} \ln (T^2 + X^2)\,, \quad
		\phi \rightarrow \frac{1}{2i} \ln \left( \frac{T + iX}{T - iX} \right)\,.
		\label{eq:dscyl-to-plane}
	\end{align}
	and the resulting conformal flat metric is 
	\begin{equation}
		\label{adsdstoflat}
		ds^2=\lds^2\sin^2w_c\frac{4\left(dT^2+dX^2\right)}{\left(1+T^2+X^2\right)^2}\,.
	\end{equation}
	The two endpoints of the subregion $A$ in the $(T,X)$ coordinates are
	\begin{equation}
		(T_1,X_1)=(e^{\tilde{t}_0}\cos \phi_a,-e^{\tilde{t}_0}\sin \phi_a),\quad (T_2,X_2)=(e^{\tilde{t}_0}\cos \phi_a,e^{\tilde{t}_0}\sin \phi_a)\,.
	\end{equation}
	Using the CHM map,
	\begin{align}
		\label{CHM in adsds}
		T\rightarrow\frac{e^{\tilde{t}_0}\sin \phi_a\sin\Theta \sin \Phi}{1+\sin\Theta\cos\Phi}+e^{\tilde{t}_0}\cos \phi_a\,,\quad
		X\rightarrow\frac{e^{\tilde{t}_0}\sin \phi_a\cos\Theta}{1+\sin\Theta\cos\Phi}\,,
	\end{align}
	this complex plane can be mapped onto the sphere, yielding the conformal spherical metric
	\begin{equation}
		ds^2= \frac{4e^{2\tilde{t}_0}\lds^2\sin^2w_c\sin^2\phi_a\left(d\Theta^2+\sin^2\Theta d\Phi^2\right)} {\left(1+\sin\Theta\cos\Phi+e^{2\tilde{t}_0}\sin\Theta\cos(\Phi-2\phi_a)+e^{2\tilde{t}_0}\right)^2}\,.
	\end{equation}
	It is easy to verify that the conformal factor at the endpoints $\Theta = 0, \pi$ is the same constant, which can be absorbed into $r_\eff$. Therefore, the effective radius is $r_{\text{eff}}=\lds\sin w_c\cos t_0\sin \phi_a$. Finally, in Lorentz signature, the effective radius is given by 
	\begin{equation}
		r_{\text{eff}}=\lds\sin w_c\cosh t_0\sin \phi_a	\,.
	\end{equation}
	As mentioned in Section~\ref{The dS case}, we require that the effective spherical radius be less than $\lds$, i.e.,
	\begin{equation}
		\label{require}
		\sin w_c\cosh t_0\sin\phi_a<1.
	\end{equation}
	In fact, this ensures that the geodesic length connecting the endpoints of the selected subregion in dS spacetime is real~\cite{Geng:2020kxh, Dong:2018cuv}. Finally, we obtain the entanglement entropy of subregion $A$ as
	\begin{equation}
		\label{dsdsfield}
		S_A
		=\frac{c_\ds}{3}\tan^{-1}\left(\frac{1}{\sqrt{\frac{c_\ds\lambda_\ds }{12r_\eff^2}-1}}\right)
		=\frac{c_\ds}{3}\tan^{-1}\left(\frac{1}{\sqrt{\frac{1}{(\sin w_c \cosh t_0\sin\phi_a)^2}-1}}\right)\,.
	\end{equation}
	For the specific case $t=0$, 
	the entanglement entropy at $t_0=0$ becomes
	\begin{equation}
		S_A=\frac{c_\ds}{3}\cot^{-1}\left(\sqrt{\cot^2\phi_a+\frac{\cot^2w_c}{\sin^2\phi_a}}\right)\,,
		\label{eq:SA_dsds}
	\end{equation}
	which is consistent with the holographic calculation of the area of the RT surface in ref.~\cite{Geng:2019ruz}. For the case where the subsystem is defined as $\phi\in(-\frac{\pi}{2},\frac{\pi}{2})$, the entanglement entropy is given by $S_A=\frac{c_\ds}{3}w_c$, which agrees with the result in ref.~\cite{Gorbenko:2018oov}. The subsequent task is to determine whether the system's entanglement entropy satisfies both SSA and BSSA. We will discuss the two cases of $t_0=0$ and $t_0\neq 0$ separately.

		\subsubsection{The case of $t_0=0$}
		
\begin{figure}[H]
			\centering
			\subfigure[The SSA is satisfied.]{
				\label{EDSS}
				\includegraphics[width=0.41\textwidth]{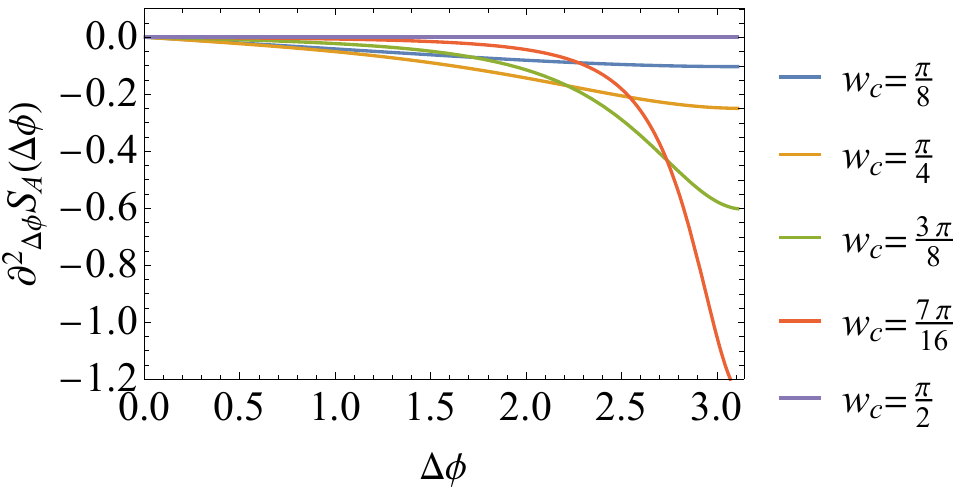}}
			\hspace{1in}
			\subfigure[The BSSA is violated.]{
				\label{EDBSSA}
				\includegraphics[width=0.39\textwidth]{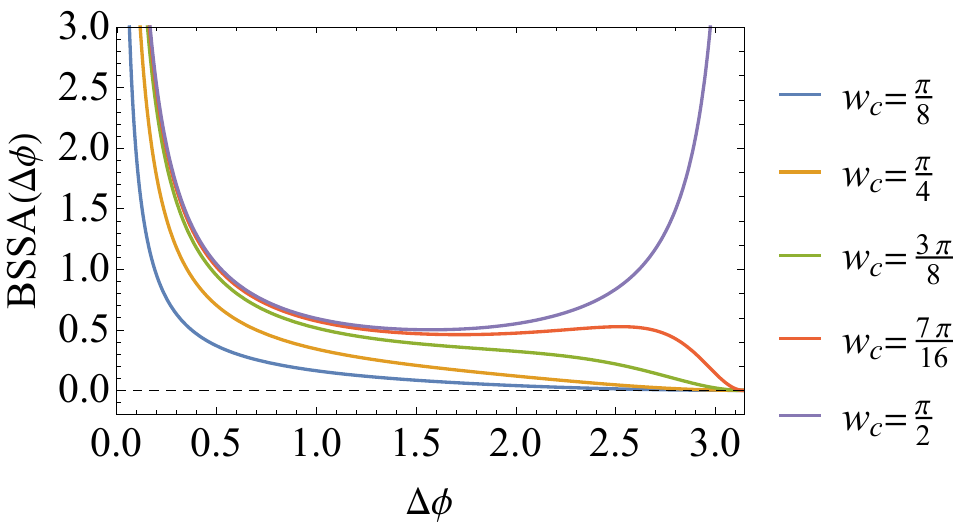}}
			\caption{The SSA and BSSA in dS/dS holography at $t_0=0$. \ref{EDSS} The SSA is satisfied at $t_0=0$. When $w_c=\frac{\pi}{2}$, the SSA is saturated. \ref{EDBSSA} The BSSA is violated at $t_0=0$. When $w_c=\frac{\pi}{2}$, the violation of the BSSA inequality is obvious.}
			\label{EDSS12}
		\end{figure}

		The second-order derivative of the entanglement entropy with respect to the subsystem scale $\Delta\phi$ ($\Delta\phi=2\phi_a$) is
		\begin{equation}
			\partial^2_{\Delta\phi}S_A(\Delta\phi)=-\frac{c_{dS}}{12\sin^2\frac{\Delta\phi}{2}}\frac{\frac{1}{\sin^2w_c}-1}{\left(\frac{1}{\sin^2\frac{\Delta\phi}{2}\sin^2w_c}-1\right)^{3/2}}<0.
		\end{equation}
		It can be observed that the SSA is manifestly satisfied, as illustrated in figure~\ref{EDSS}. However, the BSSA is violated in this case. Substituting Eq.~\eqref{eq:SA_dsds} into the left-hand side of the BSSA~\eqref{dsdsBSSA}, we obtain
		\begin{equation}
			\text{BSSA}(\Delta\phi) = \frac{c_{dS}}{48} \frac{\sin^2 \Delta \phi}{\sin^6 \frac{\Delta \phi}{2} \sin^2 w_c \left(\frac{1}{\sin^2 \frac{\Delta \phi}{2} \sin^2 w_c} - 1 \right)^{\frac{3}{2}}}>0,
		\end{equation}
		where $\text{BSSA}(\Delta\phi)$ represents $S^{\prime\prime}(\Delta\phi)+\frac{1}{\sin\Delta\phi}S^{\prime}(\Delta\phi)$. This indicates that the dual field theory continues to exhibit non-locality. Notably, at the ultraviolet boundary where $w_c = \frac{\pi}{2}$, there is a significant violation of the BSSA\footnote{In the $T\bar{T}$-deformed AdS/CFT, the BSSA violation also occurs due to the influence of the $T\bar{T}$ deformation~\cite{Aharony:2023dod}. However, as the finite truncation $r_c$ approaches the boundary, the BSSA tends to saturate, which contrasts with the behavior in the dS spacetime background.}, as shown by the purple line in figure~\ref{EDBSSA}.
		\subsubsection{The case of $t_0\neq 0$}
		In the case of $t_0=0$, $\sin w_c\cosh t_0<1$, but for $t_0\neq 0$, the value of $\sin w_c\cosh t_0$ can be either greater than 1 or less than 1. Since the effective spherical radius $r_{\text{eff}}=\ell_{dS}\sin w_c\cosh t_0\sin \frac{\Delta\phi}{2}$ is required to be less than $\ell_{dS}$, we must consider both cases, $\sin w_c\cosh t_0 < 1$ and $\sin w_c\cosh t_0 > 1$, separately.
		
		\begin{figure}[H]
			\centering
			\subfigure[The SSA is satisfied.]{
				\label{EDSS1}
				\includegraphics[width=0.39\textwidth]{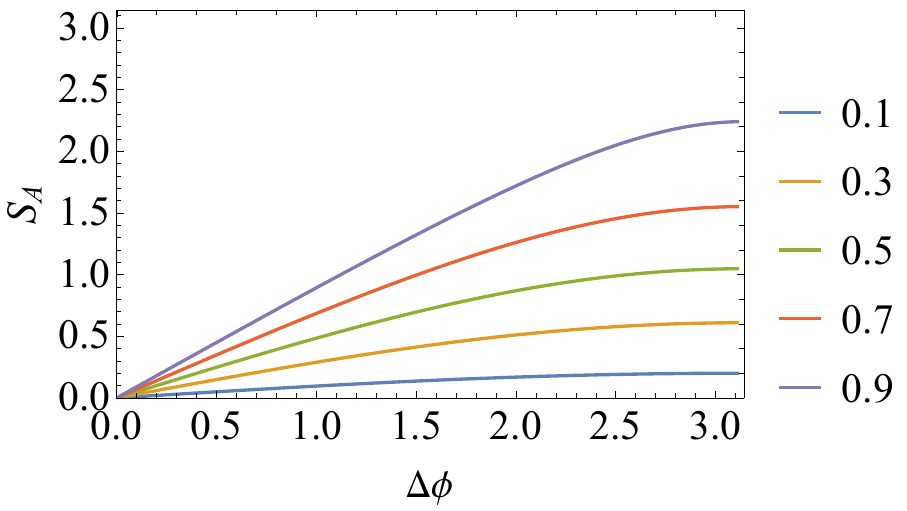}}
			\hspace{1in}
			\subfigure[The BSSA is violated.]{
				\label{EDBSSA1}
				\includegraphics[width=0.41\textwidth]{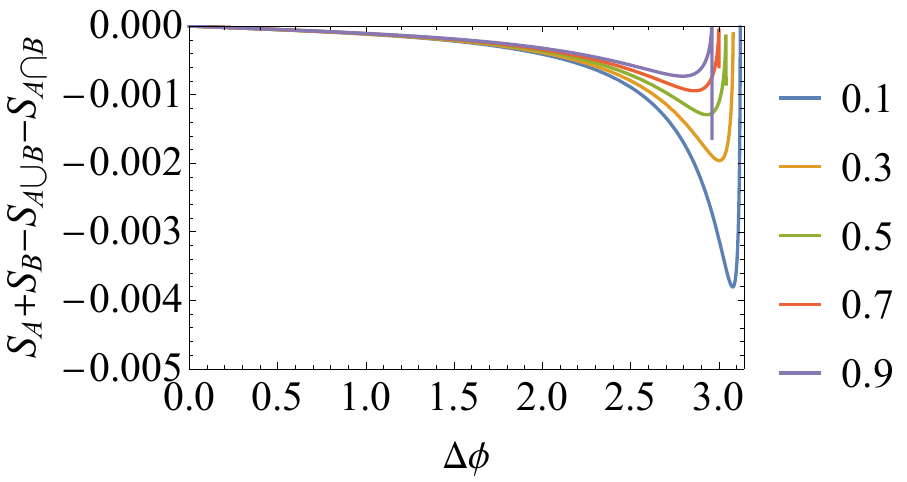}}
			\caption{The SSA and BSSA in dS/dS holography at $\sin w_c\cosh t_0 < 1$. \ref{EDSS1} The SSA is satisfied at $\sin w_c\cosh t_0<1$. The entanglement entropy, as a function of the subinterval scale $\Delta\phi$, is a concave function, ensuring that subadditivity is always satisfied. \ref{EDBSSA1} The BSSA is violated at $\sin w_c\cosh t_0<1$. To maintain $\sin w_c\cosh t_0<1$, we set $t_0$ to 0.01, 0.03, 0.05, 0.07, and 0.09, respectively, and take $\delta\phi=0.01\Delta\phi$.}
			\label{tneq01}
		\end{figure}
		When $\sin w_c\cosh t_0 < 1$, the condition $r_{\text{eff}} < \ell_{dS}$ is always satisfied for any $\Delta\phi$. Since $t\neq 0$, we directly use the entanglement entropy to verify whether SSA and BSSA are satisfied. As shown in figure~\ref{EDSS1}, when $\sin w_c\cosh t_0 < 1$, the entanglement entropy is a concave function, and the concavity of the function indicates that $\partial^2_{\Delta\phi}S(\Delta\phi,w_c,t_0)<0$, confirming that the SSA is satisfied. However, upon considering the boost transformation, we find that the BSSA is consistently violated, as shown in Figure~\ref{EDBSSA1}.
		\begin{figure}[H]
			\centering
			\subfigure[The SSA is violated.]{
				\label{EDSS2}
				\includegraphics[width=0.39\textwidth]{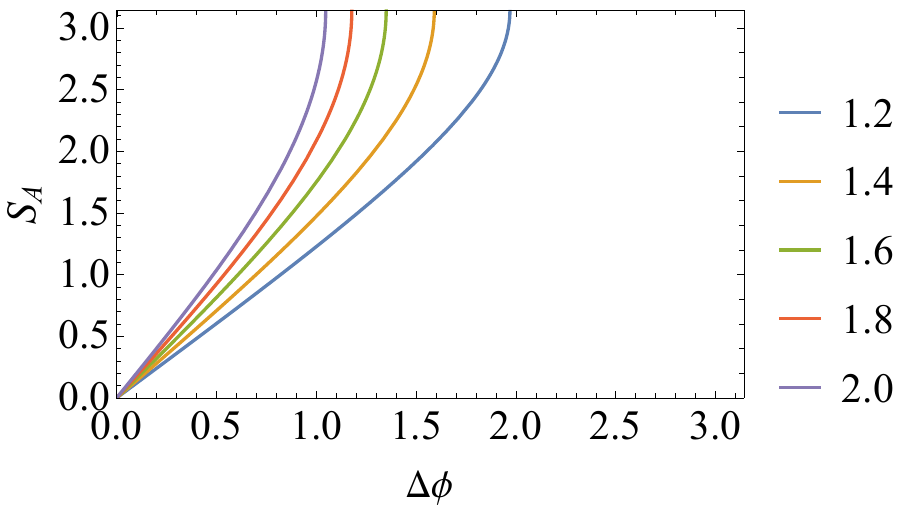}}
			\hspace{1in}
			\subfigure[The BSSA is violated.]{
				\label{EDBSSA2}
				\includegraphics[width=0.41\textwidth]{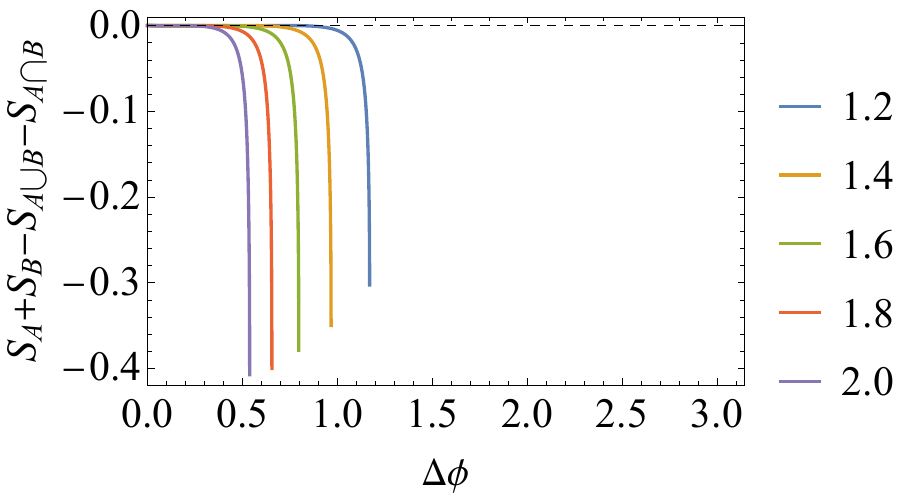}}
			\caption{The SSA and BSSA in dS/dS holography at $\sin w_c\cosh t_0 > 1$. \ref{EDSS2} The SSA is  violated at $\sin w_c\cosh t_0>1$. The entanglement entropy is a convex function, ensuring that subadditivity is always violated. \ref{EDBSSA2} The BSSA is violated at $\sin w_c\cosh t_0>1$. We set $t_0$ to 1.2, 1.4, 1.6, 1.8, and 2.0, respectively, and take $\delta\phi=0.1\Delta\phi$.}
			\label{tneq02}
		\end{figure}
		When $\sin w_c\cosh t_0 > 1$, to ensure that the effective spherical radius is less than $\lds$, the value of $\Delta\phi$ has a maximum value $\Delta\phi_{\text{max}}$ such that $\sin w_c\cosh t_0\sin\frac{\Delta\phi_{\text{max}}}{2}=1$. As shown in figure~\ref{EDSS2}, in the interval $[0, \Delta\phi_{\text{max}}]$, the entanglement entropy as a function of $\Delta\phi$ is consistently a convex function, which ensures that the subadditivity is always violated. Similarly, considering the boost transformation, it can be observed that the BSSA is still violated when $\sin w_c\cosh t_0$ is greater than 1, as shown in figure~\ref{EDBSSA2}.
		
		\begin{figure}[H]
			\centering
			\includegraphics[scale=0.8]{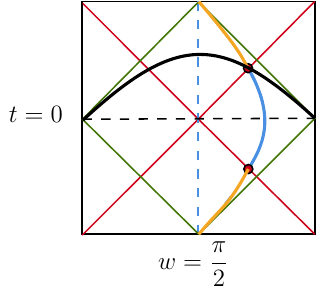}
			\caption{The Penrose diagram of dS/dS holography. The black and blue dashed lines represent the codimension-one hypersurfaces at $t = 0$ and $w = \frac{\pi}{2}$, respectively. The black curve represents the spacelike hypersurface at $t = t_0$, and the orange and blue curves represent the dS slices at $w = w_c$, where the dual field theory is located. The intersection of these two hypersurfaces corresponds to the critical case that saturates the SSA, where $\sin w_c \cosh t_0 = 1$.}
			\label{dsdsSSABSSA}
		\end{figure}
		For the two cases described above, the critical positions are illustrated in figure~\ref{dsdsSSABSSA}. The orange region corresponds to $\sin w_c\cosh t_0 > 1$, where the dual field theory is non-local. The blue region corresponds to $\sin w_c\cosh t_0 < 1$, where the dual field theory satisfies the SSA, but it still violates the BSSA. This demonstrates that the dual field theory still exhibits non-locality in the blue region in figure~\ref{dsdsSSABSSA}.

	\subsection{The half-dS holography}
	\noindent
	For the half-dS holography, the induced metric on the boundary is given by
	\begin{equation}
		ds^2 = \lds^2 (-dt^2 + \cosh^2 t \sin^2 \theta_0 d\phi^2).
	\end{equation}
	By rescaling the coordinate $t$ in a manner analogous to that discussed in Section \ref{The result of dS/dS}, we obtain the conformal cylindrical metric
	\begin{equation}
		ds^2 = \lds^2 \frac{\sin^2 \theta_0}{\cosh^2 \left( \tilde{t} \sin \theta_0 \right)} \left( d\tilde{t}^2 + d\phi^2 \right),
	\end{equation}
	where the coordinate transformation between $\tilde{t}$ and $t$ is $\tilde{t} = \frac{1}{\sin \theta_0} \ln \frac{\cot \frac{t}{2} + 1}{\cot \frac{t}{2} - 1}$.
	Mapping from the cylinder to the plane as in eq.~\eqref{eq:dscyl-to-plane}, the metric transforms into a conformally flat form
	\begin{equation}
		ds^2 = \frac{\lds^2 \sin^2 \theta_0}{\cosh^2 \left( \ln (\sqrt{T^2 + X^2}) \sin \theta_0 \right)} \frac{dT^2 + dX^2}{T^2 + X^2}.
	\end{equation}
	
\begin{figure}[H]
			\centering
			\includegraphics[scale=0.5]{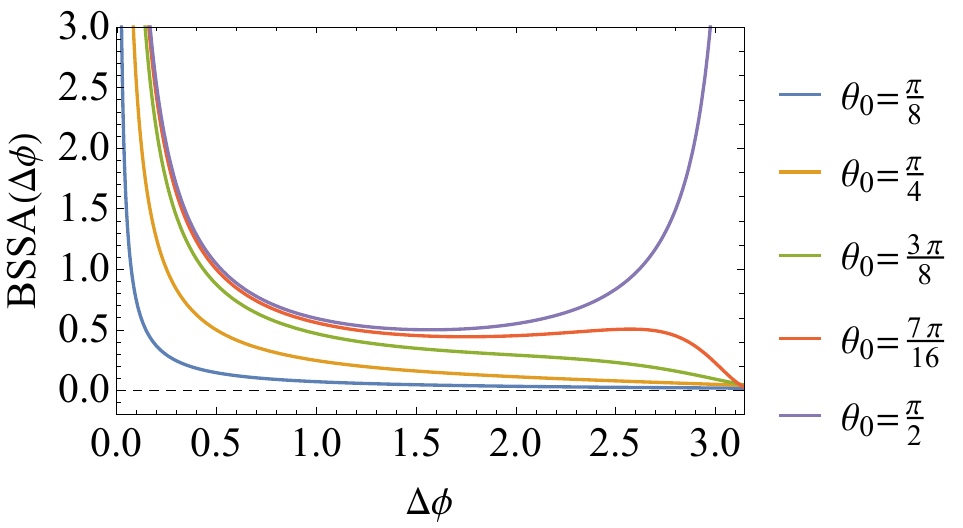}
			\caption{The boosted strong subadditivity is violated in half-dS holography.}
			\label{EDBSSAhds}
		\end{figure}

\begin{figure}[H]
			\centering
			\includegraphics[scale=0.8]{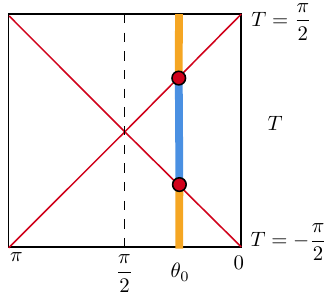}
			\caption{Half-de Sitter holography. The dual field theory resides on a timelike boundary at $\theta = \theta_0$. The blue intervals indicate $\cosh t \leq \frac{1}{\sin \theta_0}$, i.e., $\theta_0 - \frac{\pi}{2} \leq T \leq \frac{\pi}{2} - \theta_0$.}
			\label{dsstatic2}
		\end{figure}

	Next, we calculate the entanglement entropy for the subregion interval $(-\phi_a, \phi_a)$ at $t = t_0$. After the mapping, the coordinates of the subregion endpoints are
	\begin{equation}
		(T_1, X_1) = \left( e^{\tilde{t}_0} \cos \phi_a, -e^{\tilde{t}_0} \sin \phi_a \right), \quad (T_2, X_2) = \left( e^{\tilde{t}_0} \cos \phi_a, e^{\tilde{t}_0} \sin \phi_a \right).
	\end{equation}
	Using the CHM mapping,
	\begin{align}
		T\rightarrow\frac{e^{\tilde{t}_0}\sin \phi_a\sin\Theta \sin \Phi}{1+\sin\Theta\cos\Phi}+e^{\tilde{t}_0}\cos \phi_a\,,\quad
		X\rightarrow\frac{e^{\tilde{t}_0}\sin \phi_a\cos\Theta}{1+\sin\Theta\cos\Phi}\,,
	\end{align}
	we obtain the conformal spherical metric
	\begin{equation}
		ds^2 = \frac{2 \Omega^2 \lds^2 \sin^2 \theta_0 \sin^2 \phi_a \left( d\Theta^2 + \sin^2 \Theta d\Phi^2 \right)}{(1 + \sin \Theta \cos \Phi)(2 + \sin (\Theta + \Phi - 2 \phi_a) + \sin (\Theta - \Phi + 2 \phi_a))},
	\end{equation}
	where 
	\begin{equation}
		\Omega^{-1} = \cosh \left( \frac{1}{2} \ln \left( \frac{e^{2\tilde{t}_0} \left( 2 + \sin (\Theta + \Phi - 2 \phi_a) + \sin (\Theta - \Phi + 2 \phi_a) \right)}{2 + 2 \sin \Theta \cos \Phi} \right) \sin \theta_0 \right).
	\end{equation}
	At the endpoints ($\Theta = 0, \pi$), the conformal factor becomes $\cosh^{-1} \tilde{t}_0$. Thus, the effective radius of the sphere in the Lorentzian signature is
	\begin{equation}
		r_{\text{eff}} = \lds \cosh t_0 \sin \theta_0 \sin \phi_a.
	\end{equation}
		In order to ensure the existence of spacelike geodesics, the ref.~\cite{Kawamoto:2023nki} requires~\footnote{Equation (3.6) of ref.~\cite{Kawamoto:2023nki} with the substitution $\phi_1-\phi_2=2\phi_a=\Delta\phi$.}
		\begin{equation}
			\cos^2\theta_0+\sin^2\theta_0\cos(\phi_1-\phi_2)\leq 1-\frac{2}{\cosh^2t_0}.
		\end{equation}
		After simplification and deformation, it can be observed that this ensures that the effective radius of the sphere, $r_{\text{eff}} = \lds \cosh t_0 \sin \theta_0 \sin \frac{\Delta\phi}{2}$, is less than or equal to $\lds$. Therefore, the entanglement entropy is
	\begin{equation}
		\label{hadseeresult}
		S = \frac{c_\ds}{3} \tan^{-1} \left( \frac{1}{\sqrt{\frac{c_\ds \lambda_\ds }{12 r_{\text{eff}}^2} - 1}} \right)
		= \frac{c_\ds}{3} \tan^{-1} \left( \frac{1}{\sqrt{\frac{1}{(\cosh t_0 \sin \theta_0 \sin \phi_a)^2} - 1}} \right).
	\end{equation}
	In the case of $t_0 = 0$ and $\theta_0 = \frac{\pi}{2}$, the radius in the spherical partition function is $r_{\text{eff}} = \lds \sin \frac{\Delta\phi}{2}$, where $\Delta\phi = |-\phi_a - \phi_a| = 2\phi_a$. The entanglement entropy is
	\begin{equation}
		S = \frac{c_\ds}{3} \tan^{-1} \left( \frac{1}{\sqrt{\frac{c_\ds \lambda_\ds }{12 r_{\text{eff}}^2} - 1}} \right)
		= \frac{c_\ds}{3} \tan^{-1} \left( \frac{1}{\sqrt{\frac{1}{\sin^2 \frac{\Delta\phi}{2}} - 1}} \right)
		= \frac{c_\ds}{6} \Delta\phi.
	\end{equation}
	By replacing the field theory parameter $c_\ds$ with the gravitational parameter $\frac{3 \lds}{2 G}$, we find that the above result matches the holographic calculation presented in ref.~\cite{Kawamoto:2023nki}, confirming the validity of half-dS holography. By comparing Eqs.~\eqref{dsdsfield} and~\eqref{hadseeresult}, we find that the entanglement entropy results for spatial intervals, calculated independently in dS/dS holography and half-dS holography, reveal a close analogy between the two frameworks. Therefore, the discussion of SSA aligns with that in dS/dS holography. However, due to the difference in boundary-induced metrics, the BSSA receives corresponding corrections. In half-dS holography, the infinitesimal form of the BSSA is
		\begin{equation}
			\text{BSSA}(\Delta\tilde{\phi})
			=\sin^2\theta_0\left(\partial^2_{\Delta\tilde{\phi}}S(\Delta\tilde{\phi})+\frac{1}{\sin\Delta\tilde{\phi}}\partial_{\Delta\tilde{\phi}}S(\Delta\tilde{\phi})\right)\leq 0,
		\end{equation}
		where $\Delta\tilde{\phi}=\sin\theta_0\Delta\phi$. Substituting the entanglement entropy result into the above expression, we obtain
		\begin{equation}
			\text{BSSA}(\Delta\phi)=\frac{c_{dS}}{48}\frac{\sin^2(\Delta\phi\sin\theta_0)}{\sin^6(\frac{1}{2}\Delta\phi\sin\theta_0)(\frac{1}{\sin^2\theta_0\sin^2(\frac{1}{2}\Delta\phi\sin\theta_0)}-1)^{\frac{3}{2}}}>0.
		\end{equation}
		The result is illustrated in figure~\ref{EDBSSAhds}. Compared with figure~\ref{EDBSSA}, it is evident that figure~\ref{EDBSSAhds} exhibits only slight differences. The primary behavior of the function remains nearly identical, and a clear deviation in behavior is observed at $\theta_0=\frac{\pi}{2}$, indicating that the boundary dual field theory in half-dS holography continues to exhibit non-locality.		
		\noindent
		Similarly, there is a critical position $\sin\theta_0\cosh t_0=1$ in half-dS holography, as indicated by the red dot in figure~\ref{dsstatic2}. As depicted in figure~\ref{dsstatic2}, the orange line segment represents $\sin\theta_0\cosh t_0>1$, where the SSA is violated, while the blue line segment represents $\sin\theta_0\cosh t_0<1$, where the SSA is satisfied, but the BSSA is still violated.
		We have computed the entanglement entropy of subsystem $A$ using the generalized spherical replica method within the dS/dS holography and half-dS holography frameworks. By verifying the SSA and BSSA, we have found that the entanglement entropy exhibits similar behavior in these two holographic models and that the boundary dual field theory exhibits non-locality. Comparing figures~\ref{dsdsSSABSSA} and~\ref{dsstatic2}, we observe the same critical point, namely, when the region where the dual field theory is located crosses the cosmological horizon in three-dimensional dS$_3$ spacetime, the entanglement entropy violates the SSA, and even within the cosmological horizon, the entanglement entropy still violates the BSSA.

	\section{Conclusion and Discussion}
	\label{Conclusion and Discussion}
	\noindent 
		This paper systematically investigated the impact of the $T\bar{T}$ deformation on various holographic dualities, including AdS/CFT, dS/CFT, dS/dS, and half-dS holography, by analyzing the entanglement entropy of the single general spatial subinterval in the boundary dual field theory. When the boundary is spherical, and the subsystem is a semicircular arc, the entanglement entropy can be derived from the spherical partition function on the $n$-replica manifold. For general boundary manifolds, we first mapped it onto a conformal spherical manifold via CHM mapping. Then, we calculated the entanglement entropy using the corresponding effective spherical radius, and derived the correction after $T\bar{T}$ deformation in both the dS Poincaré spacetime and global dS spacetime. In addition, we found that the dual field theory exhibits notable similarities in properties between the dS/dS holography and half-dS holography models. Specifically, the dual field theory in both cases exhibited non-local characteristics. When the dual field theory lies within the cosmological horizon, it satisfies strong subadditivity but violates boosted strong subadditivity. Conversely, when the dual field theory lies outside the cosmological horizon, both strong subadditivity and boosted strong subadditivity are violated.
	\acknowledgments
	We would like to thank Shan-Ming Ruan, Yuan Sun for reading our draft and providing comments. We also thank Yang He, and Shan-Ping Wu for their helpful discussion. This work was supported by the National Natural Science Foundation of China (Grants No. 12475056, No. 12235016, No. 12075101, No. 12475053, and No. 12247101), the 111 Project under (Grant No. B20063), the Major Science and Technology Projects of Gansu Province, and Lanzhou City's scientific research funding subsidy to Lanzhou University. S.H. would appreciate the financial support from the Fundamental Research Funds for the Central Universities and Max Planck Partner Group and the Natural Science Foundation of China (NSFC) Grants No.~12075101 and No.~12235016. L.Z. is supported by the Science and Technology
	Development Plan Project of Jilin Province, China (No. 20240101326JC).
	
	\section{Data Availability Statement}	
	The data that support the findings of this article are openly available~\cite{ChangHe2025Data}

	\appendix
		\section{Integration over the endpoints of the entanglement region}
		\label{endpoints}
		\noindent For a two-dimensional spherical manifold $\mathcal{M}$, on the slice where $\phi = \text{constant}$, we select $\theta \in [0, \pi]$ as the subregion $A$, the entanglement entropy $S_A$ can be obtained through the $C$-function. Differentiating $S_A$ with respect to $r$, we have
		\begin{align}
			r \frac{d}{dr} S_A(r)
			&= r \frac{d}{dr} \lim_{n \rightarrow 1} (1 - n \partial_n) \ln Z_n
			= \lim_{n \rightarrow 1} (1 - n \partial_n) r \frac{d}{dr} \ln Z_n \nn \\
			&= -\lim_{n \rightarrow 1} (1 - n \partial_n) 2\pi n r^2 \int_0^\pi \sin\theta d\theta \, \text{tr}\, T \nn \\
			&= 2\pi r^2 \lim_{n \rightarrow 1} n^2 \int_0^\pi \sin\theta d\theta \, \partial_n \text{tr}\, T.
			\label{eq:d-Cfunction}
		\end{align}
		Solving the trace flow and conservation equations without assuming spherical symmetry must determine the energy-momentum tensor $T_{ab}$. For the $n$-replica sphere \eqref{eq:metric-n-sphere}, the solution for the energy-momentum tensor is known~\cite{Donnelly:2018bef} and is given by
		\begin{align}
			T^{\theta}_\theta
			&= \frac{1}{\pi \lambda} \left( 1 - \sqrt{1 + \frac{\lambda c}{12 r^2} + \frac{\lambda c}{12 r^2} \left(\frac{1}{n^2} - 1\right) \frac{1}{\sin^2\theta}} \right), \nn \\
			T^{\phi}_\phi
			&= \frac{1}{\pi \lambda} \left( 1 - \frac{1 + \frac{\lambda c}{12 r^2}}{\sqrt{1 + \frac{\lambda c}{12 r^2} + \frac{\lambda c}{12 r^2} \left(\frac{1}{n^2} - 1\right) \frac{1}{\sin^2\theta}}} \right).
			\label{eq:T-nsphere}
		\end{align}
		Substituting this into the integral in eq.~\eqref{eq:d-Cfunction}, in the limit $n \rightarrow 1$, the integrand becomes
		\begin{align}
			n \partial_n \text{tr}\, T = \frac{c}{3} \frac{\lambda c}{48\pi} \frac{1-n^2}{n^4 r^4 \sin^4 \theta} \frac{1}{\left[1 + \frac{\lambda c}{12 r^2} \left(1 + \frac{2(1 - n)}{\sin^2 \theta}\right)\right]^{3/2}},
		\end{align}
		which tends to zero for $\theta \neq 0, \pi$. However, near the branch points, the integrand has a non-zero contribution. Thus, the main contribution to the integral in eq.~\eqref{eq:d-Cfunction} comes from small neighborhoods near the branch points
		\begin{align}
			r S'(r) = 2\pi r^2 \lim_{n \rightarrow 1} \left( \int_{\theta \in B(0, \epsilon)} + \int_{\theta \in B(\pi, \epsilon)} \right) \sin\theta d\theta \, \partial_n \text{tr}\, T,
			\label{eq:d-Cfunction-endpoint}
		\end{align}
		where $B(p, \epsilon)$ is a small neighborhood centered at $p$ with radius $\epsilon$. 
		\subsection{plane}
		For a plane manifold $\mathcal{M}$, we first map $\mathcal{M}$ to $\mathbb{S}^2$ by CHM map, ensuring that the image of the subsystem $A$ under this mapping is a meridian on $\mathbb{S}^2$. The mapped metric is conformally equivalent to the sphere, denoted as $\Omega \mathbb{S}^2$. After applying the $n$-replica trick, we assume the metric takes the form\footnote{We use capital Greek letters to represent the angular coordinates of the conformal sphere $\Omega \mathbb{S}^2$}
		\begin{align}
			ds^2 = r^2 \Omega^2(\Theta, \Phi) \left(d\Theta^2 + n^2 \sin^2\Theta \, d\Phi^2\right),
			\label{eq:metric-conformalsphere}
		\end{align}
		where the conformal factor $\Omega(\Theta, \Phi)$ is constant at the branch points $\Theta = 0$ and $\Theta = \pi$, denoted as $\Omega_0$ and $\Omega_\pi$, respectively. For this metric, eqs.~\eqref{eq:dlnz/dr} and \eqref{eq:dlnzn/dn} remain valid.
		Thus, we have
		\begin{align}
			r \frac{d}{dr} S(r)
			&= -\lim_{n \rightarrow 1} (1 - n \partial_n) n r^2 \int_{\Omega\mathbb{S}_n^2} \Omega^2(\Theta, \Phi) \sin\Theta \, d\Theta d\Phi \, \text{tr}\, T \nn \\
			&= r^2 \lim_{n \rightarrow 1} n^2 \int_{\Omega\mathbb{S}_n^2} \Omega^2(\Theta, \Phi) \sin\Theta \, d\Theta d\Phi \, \partial_n \text{tr}\, T.
			\label{eq:dlnzn/dn-conformalsphere}
		\end{align}
		Since the conformal factor $\Omega(\Theta, \Phi)$ depends on the azimuthal angle $\Phi$, $T_{ab}$ cannot be assumed to depend solely on $\Theta$. Therefore, the $\Phi$ coordinate is not integrated. As discussed in ref.~\cite{Lewkowycz:2019xse}, for the partition function on the sphere~\eqref{eq:d-Cfunction}, in the limit $n \rightarrow 1$, the main contribution to the integral~\eqref{eq:d-Cfunction} comes from a small neighborhood near the branch points, with radius $\epsilon \sim \sqrt{n - 1}$. This property is expected to hold for any manifold $\mathcal{M}$, so the integral~\eqref{eq:dlnzn/dn-conformalsphere} is also performed only in small neighborhoods near the branch points. In these regions, the conformal factor $\Omega$ is constant, and the metric~\eqref{eq:metric-conformalsphere} can be approximated by the $n$-replica sphere with radius $r \Omega$. As a result, the conformal transformation can be regarded as an identity transformation\footnote{ In standard CFT calculations, a conformal transformation is often applied to map $\Omega\mathbb{S}^2$ to the spherical metric, and conformal symmetry ensures that the results for $\Omega \mathbb{S}^2$ and $\mathbb{S}^2$ differ only by a function of the conformal factor. However, for the $T\bar{T}$-deformed field theory, conformal symmetry is absent, and the relationship between quantities before and after the conformal transformation is not determined.}. Therefore, partition functions on any manifold $\mathcal{M}$ can be calculated using the results on $\mathbb{S}^2$ for non-conformal field theories. Finally, we can get
		\begin{align}
			r S'(r)
			= \lim_{n \rightarrow 1} 2\pi (r \Omega)^2 \left(\int_0^{\Theta_0} + \int_{\pi - \Theta_0}^\pi \right) \sin\Theta \, d\Theta \, \partial_n \text{tr}\, T,
			\label{eq:dC-conformalsphere}
		\end{align}
		where $\text{tr}\, T$ corresponds to the solution on the sphere with radius $r \Omega$. 	When Poincaré coordinates are chosen in EAdS$_3$, the holographic boundary corresponds to a two-dimensional complex plane with the metric
		\begin{align}
			\label{eq:poincare}
			ds_2^2 = \frac{\lads^2}{z_c^2}(dT^2 + dX^2).
		\end{align}
		The subregion $A$ is defined by $T = 0$ and $X \in [-X_a, X_a]$. To compute the entanglement entropy for $A$, the complex plane is mapped onto a two-dimensional sphere such that the image of $A$ corresponds to the meridian on the sphere at $\Phi = 0$. This mapping, known as the CHM mapping, is given by
		\begin{align}
			X = \frac{X_a \cos\Theta}{1 + \sin\Theta \cos\Phi}, \quad
			T = \frac{X_a \sin\Theta \sin\Phi}{1 + \sin\Theta \cos\Phi}.
			\label{eq:CHM}
		\end{align}
		Under this transformation, the boundary plane in $(\Theta, \Phi)$ coordinates has the metric
		\begin{align}
			ds_2^2 = \frac{\lads^2 X_a^2}{z_c^2 (1 + \cos\Phi \sin\Theta)^2}(d\Theta^2 + \sin^2\Theta \, d\Phi^2),
			\label{eq:pl-CHM}
		\end{align}
		which is conformally equivalent to a sphere with an effective radius $r_\mathrm{eff} = \frac{\lads X_a}{z_c}$, and the conformal factor $\Omega(\Theta, \Phi) = (1 + \cos\Phi \sin\Theta)^{-1}$. As previously noted, when computing the partition function on a general manifold $\mathcal{M}$, the primary contribution to the integral~\eqref{eq:d-Cfunction} comes from an infinitesimal neighborhood near the branch points. In these regions, the conformal factor $\Omega(\Theta, \Phi)$ can be approximated as 1, implying that the metric~\eqref{eq:pl-CHM} effectively describes a sphere with radius
		\begin{equation}
			r_\mathrm{eff} = \frac{\lads X_a}{z_c}\,.
		\end{equation}
		Thus, within the infinitesimal neighborhoods around the endpoints, transforming~\eqref{eq:poincare} to the sphere involves only a conformal transformation with a conformal factor of 1, effectively an identity transformation. Consequently, the partition function remains the same as that of the sphere, with the radius $r$ replaced by the effective radius $r_\mathrm{eff}$. Therefore, the solution for the energy-momentum tensor on the $n$-replica sphere is
		\begin{align}
			T^{\Theta}_\Theta
			&=\frac{1}{\pi\lambda_\ads}
			\left(1-\sqrt{1+\frac{\lambda_\ads c_\ads}{12r_\eff^2}+\frac{\lambda_\ads c_\ads}{12r_\eff^2}\left(\frac{1}{n^2}-1\right)\frac{1}{\sin^2\Theta}}\right),\nn\\
			T^{\Phi}_\Phi
			&= \frac{1}{\pi\lambda_\ads}
			\left(1-\frac{1+\frac{\lambda_\ads c_\ads}{12r_\eff^2}}{\sqrt{1+\frac{\lambda_\ads c_\ads}{12 r_\eff^2}+\frac{\lambda_\ads c_\ads}{12r_\eff^2}\left(\frac{1}{n^2}-1\right)\frac{1}{\sin^2\Theta}}} \right)\,.
			\label{eq:T-nphere-AdS}
		\end{align}
		Substituting this into eq.~\eqref{eq:d-Cfunction} and performing the integration near the endpoints yields the $C$-function as
		\begin{align}
			\label{cfunction of ads}
			C(r)=\frac{c_\ads}{3}\frac{1}{\sqrt{1+\frac{c_\ads \lambda_\ads}{12 r_\eff^2}}}\,.
		\end{align}
		So we can get the entanglement entropy of the subregion A as
		\begin{align}
			S_A=\frac{c_\ads}{3} \tanh^{-1}\left(\frac{1}{\sqrt{1 + \frac{ z_c^2}{X_a^2}}}\right)
			\simeq\frac{c_\mathrm{AdS}}{3} \log \frac{2X_a}{z_c} + \frac{c_\mathrm{AdS}}{12} \frac{z_c^2}{X_a^2}.
		\end{align}

	\section{The gravity calculation}
	
	\subsection{The dS Poincaré spacetime}
	\label{appx:RT-dS/Poincaré}
	\noindent In the dS Poincaré spacetime,
	\begin{align}
		ds^2 = \frac{\lds^2}{\eta^2}(-d\eta^2+dX^2+dT^2)\,,
	\end{align}
	the geodesic distance $D_{12}$ between two points ``1'' $(\eta,X,T)=(\eta_c,-X_a,0)$ and ``2'' $(\eta_c,-X_a,0)$ can be desolved by
	\begin{equation}
		\cos \frac{D_{12}}{\lds }=-V_1V_2+U_1U_2+Y_1Y_2+W_1W_2.
	\end{equation}
	The uppercase coordinates ``$U$, $Y$, $W$, $V$'' denote the coordinates of the embedded spacetime
	\begin{equation}
		ds^2=-dV^2+dW^2+dU^2+dY^2.
	\end{equation}
	The dS spacetime is a codimensional one hypersurface in the embedded spacetime
	\begin{equation}
		-V^2+W^2+U^2+Y^2=\lds ^2,
	\end{equation}
	and they adhere to the following transformation relations with respect to the coordinates of the dS spacetime itself
	\begin{equation}
		\begin{aligned}
			U&=\frac{\lds }{\eta}X,\quad V=\lds \sinh(\log(\frac{\lds }{\eta}))+\frac{1}{\eta}\frac{X^2+T^2}{2},\\
			Y&=\frac{\lds }{\eta}T,\quad W=\lds \cosh(\log(\frac{\lds }{\eta}))-\frac{1}{\eta}\frac{X^2+T^2}{2}.\\
		\end{aligned}
	\end{equation}
	Finally we can get that the geodesic distance is
	\begin{equation}
		D_{12}=\lds\cos^{-1}\left(\lds^2\left(1-\frac{2X_a^2}{\eta_c^2}\right) \right)\simeq-2i \lds \log\frac{2X_a}{\eta_c}+i \lds \frac{\eta_c^2}{2X_a^2}+\pi \lds .
	\end{equation}
	So the entanglement entropy of the subregion is
	\begin{equation}
		S=\frac{D_{12}}{4G}\simeq-\frac{ic_\ds}{3}\log\frac{2X_a}{\eta_c}+\frac{ic_\ds\eta_c^2}{12X_a^2}+\frac{\pi c_\ds}{6}.
	\end{equation}
	This is consistent with eq.~\eqref{eq:EE-poincare-ds}.

	\subsection{The global dS spacetime}
	\label{appx:RT-globalds}
	\noindent 
	For the dS global coordinates, the coordinate transformation in embedding spacetime is
	\begin{equation}
		\begin{aligned}
			U&=\lds \cosh t\sin\theta\cos\phi,\quad V=\lds \sinh t,\\
			Y&=\lds \cosh t\sin\theta\sin\phi,\quad W=\lds \cosh t\cos\theta.\\	
		\end{aligned}
	\end{equation}
	So the geodesic distance of two points $(t,\theta,\phi)$=$(t_1,\theta_1,\phi_1)$, $(t_2,\theta_2,\phi_2)$ in the dS global coordinates is
	\begin{equation}
		\cos \frac{D_{12}}{\lds }=\left(\cos\theta_1\cos\theta_2+\sin\theta_1\sin\theta_2\cos(\phi_1-\phi_2)\right)\cosh t_1\cosh t_2-\sinh t_1\sinh t_2.
	\end{equation}
	Now we can select the endpoints of the subsystem as $(t,\theta,\phi)$=$(t_c,\frac{\pi}{2},-\phi_a)$, $(t_c,\frac{\pi}{2},\phi_a)$, so we have
	\begin{equation}
		D_{12}=\lds \cos^{-1}\left(\cos(2\phi_a)\cosh^2 t_c-\sinh^2 t_c\right)=\lds \cos^{-1}\left(1-2\cosh^2 t_c\sin^2\phi_a\right).
	\end{equation}
	Considering that the finite cutoff $t_c$ tends to infinity, we expand the above results to the second-order 
	\begin{equation}
		D_{12}\simeq-2i\lds \log(2\cosh t_c\sin\phi_a)+\frac{i\lds }{2\cosh^2t_c\sin^2\phi_a}+\pi \lds .
	\end{equation}
	Finally, we have the entanglement entropy 
	\begin{equation}
		S=\frac{D_{12}}{4G}\simeq-\frac{ic_\ds}{3}\log(2\cosh t_c\sin\phi_a)+\frac{ic_\ds}{12\cosh^2t_c\sin^2\phi_a}+\frac{\pi c_\ds}{6}.
	\end{equation}
	We can get the same result as eq.~\eqref{eq:EE-ds-global}.
	\section{Strong subadditivity and boosted strong subadditivity}
	\label{Strong subadditivity and boosted strong subadditivity}
\begin{figure}[H]
			\centering
			\subfigure[]{
				\label{SSA}
				\includegraphics[width=0.3\textwidth]{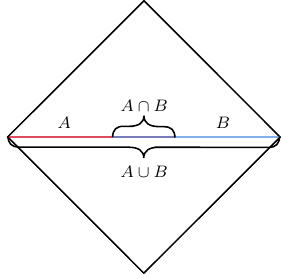}}
			\hspace{1in}
			\subfigure[]{
				\label{BSSA}
				\includegraphics[width=0.3\textwidth]{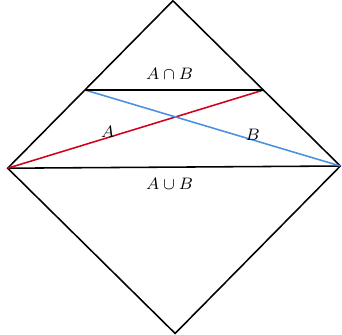}}
			\caption{\ref{SSA} Strong subadditivity. \ref{BSSA} Boosted strong subadditivity.
			}
			\label{fig:boosted strong subadditivity.}
		\end{figure}

	\noindent	For a unitary and Poincaré-invariant field theory, the strong subadditivity is given by:
		\begin{equation}
			\label{bssa}
			S(A) + S(B) - S(A \cup B) - S(A \cap B) \geq 0,
		\end{equation}
		where the subsystems $A$ and $B$ are shown in figure~\ref{SSA}. This inequality can be derived from the monotonicity of relative entropy, which comes from the locality of field theory~\cite{Witten:2018zxz}. According to the RT formula, the entanglement entropy is quantified by the area of the extremal surface, and thus the entanglement entropy can be expressed as a function of the subsystem size, such as $S(A)\equiv S(L_A)$. If we set $L_{A \cap B}=L$, $L_{A \cup B}=L+2\delta L$, the lengths of $A$ and $B$ are both $L_A=L_B=L+\delta L$. In the limit $\delta L\to 0$, the strong subadditivity can be expressed as
		\begin{equation}
			L^2 S^{\prime\prime}(L) \leq 0,
		\end{equation}
		where $L$ represents the size of the subsystems.	
		\noindent A stronger criterion for proving locality, compared to the strong subadditivity, is the boosted strong subadditivity, which can be derived by applying a boost transformation to subsystems $A$ and $B$, as shown in figure~\ref{BSSA}.  We also set $L_{A \cap B}=L$, $L_{A \cup B}=L+2\delta L$. In Minkowski spacetime, the lengths of $A$ and $B$ are both $L_A=L_B=\sqrt{L(L+2\delta L)}$. In the limit $\delta L\to 0$, the boosted strong subadditivity can be expressed as
		\begin{equation}
			S^{\prime\prime}(L) + \frac{S^{\prime}(L)}{L} \leq 0.
	\end{equation}

	\bibliographystyle{JHEP}
	
	\bibliography{ref2}

\providecommand{\href}[2]{#2}\begingroup\raggedright\begin{thebibliography}{100}

\bibitem{Smirnov:2016lqw}
F.A.~Smirnov and A.B.~Zamolodchikov, \emph{{On space of integrable quantum
  field theories}},
  \href{https://doi.org/10.1016/j.nuclphysb.2016.12.014}{\emph{Nucl. Phys. B}
  {\bfseries 915} (2017) 363}
  [\href{https://arxiv.org/abs/1608.05499}{{\ttfamily 1608.05499}}].

\bibitem{Cavaglia:2016oda}
A.~Cavagli\`a, S.~Negro, I.M.~Sz\'ecs\'enyi and R.~Tateo, \emph{{$T
  \bar{T}$-deformed 2D Quantum Field Theories}},
  \href{https://doi.org/10.1007/JHEP10(2016)112}{\emph{JHEP} {\bfseries 10}
  (2016) 112} [\href{https://arxiv.org/abs/1608.05534}{{\ttfamily
  1608.05534}}].

\bibitem{Zamolodchikov:2004ce}
A.B.~Zamolodchikov, \emph{{Expectation value of composite field T anti-T in
  two-dimensional quantum field theory}},
  \href{https://arxiv.org/abs/hep-th/0401146}{{\ttfamily hep-th/0401146}}.

\bibitem{Jiang:2019tcq}
Y.~Jiang, \emph{{Expectation value of $\mathrm{T}\overline{\mathrm{T}}$
  operator in curved spacetimes}},
  \href{https://doi.org/10.1007/JHEP02(2020)094}{\emph{JHEP} {\bfseries 02}
  (2020) 094} [\href{https://arxiv.org/abs/1903.07561}{{\ttfamily
  1903.07561}}].

\bibitem{Maldacena:1997re}
J.M.~Maldacena, \emph{{The Large N limit of superconformal field theories and
  supergravity}}, \href{https://doi.org/10.1023/A:1026654312961}{\emph{Adv.
  Theor. Math. Phys.} {\bfseries 2} (1998) 231}
  [\href{https://arxiv.org/abs/hep-th/9711200}{{\ttfamily hep-th/9711200}}].

\bibitem{Gubser:1998bc}
S.S.~Gubser, I.R.~Klebanov and A.M.~Polyakov, \emph{{Gauge theory correlators
  from noncritical string theory}},
  \href{https://doi.org/10.1016/S0370-2693(98)00377-3}{\emph{Phys. Lett. B}
  {\bfseries 428} (1998) 105}
  [\href{https://arxiv.org/abs/hep-th/9802109}{{\ttfamily hep-th/9802109}}].

\bibitem{Witten:1998qj}
E.~Witten, \emph{{Anti-de Sitter space and holography}},
  \href{https://doi.org/10.4310/ATMP.1998.v2.n2.a2}{\emph{Adv. Theor. Math.
  Phys.} {\bfseries 2} (1998) 253}
  [\href{https://arxiv.org/abs/hep-th/9802150}{{\ttfamily hep-th/9802150}}].

\bibitem{McGough:2016lol}
L.~McGough, M.~Mezei and H.~Verlinde, \emph{{Moving the CFT into the bulk with
  $ T\overline{T} $}},
  \href{https://doi.org/10.1007/JHEP04(2018)010}{\emph{JHEP} {\bfseries 04}
  (2018) 010} [\href{https://arxiv.org/abs/1611.03470}{{\ttfamily
  1611.03470}}].

\bibitem{Kraus:2018xrn}
P.~Kraus, J.~Liu and D.~Marolf, \emph{{Cutoff AdS$_{3}$ versus the $
  T\overline{T} $ deformation}},
  \href{https://doi.org/10.1007/JHEP07(2018)027}{\emph{JHEP} {\bfseries 07}
  (2018) 027} [\href{https://arxiv.org/abs/1801.02714}{{\ttfamily
  1801.02714}}].

\bibitem{He:2023hoj}
S.~He, Y.~Li, Y.-Z.~Li and Y.~Zhang, \emph{{Holographic torus correlators of
  stress tensor in AdS$_{3}$/CFT$_{2}$}},
  \href{https://doi.org/10.1007/JHEP06(2023)116}{\emph{JHEP} {\bfseries 06}
  (2023) 116} [\href{https://arxiv.org/abs/2303.13280}{{\ttfamily
  2303.13280}}].

\bibitem{He:2023knl}
S.~He, Y.-Z.~Li and Y.~Zhang, \emph{{Holographic torus correlators in AdS$_{3}$
  gravity coupled to scalar field}},
  \href{https://doi.org/10.1007/JHEP05(2024)254}{\emph{JHEP} {\bfseries 05}
  (2024) 254} [\href{https://arxiv.org/abs/2311.09636}{{\ttfamily
  2311.09636}}].

\bibitem{He:2024fdm}
S.~He, Y.~Li, Y.-Z.~Li and Y.~Zhang, \emph{{Note on holographic torus stress
  tensor correlators in AdS$_{3}$ gravity}},
  \href{https://doi.org/10.1007/JHEP09(2024)125}{\emph{JHEP} {\bfseries 09}
  (2024) 125} [\href{https://arxiv.org/abs/2405.01255}{{\ttfamily
  2405.01255}}].

\bibitem{He:2024xbi}
S.~He, Y.-Z.~Li and Y.~Xie, \emph{{Holographic stress tensor correlators on
  higher genus Riemann surfaces}},
  \href{https://doi.org/10.1007/JHEP10(2024)208}{\emph{JHEP} {\bfseries 10}
  (2024) 208} [\href{https://arxiv.org/abs/2406.04042}{{\ttfamily
  2406.04042}}].

\bibitem{Hubeny:2007xt}
V.E.~Hubeny, M.~Rangamani and T.~Takayanagi, \emph{{A Covariant holographic
  entanglement entropy proposal}},
  \href{https://doi.org/10.1088/1126-6708/2007/07/062}{\emph{JHEP} {\bfseries
  07} (2007) 062} [\href{https://arxiv.org/abs/0705.0016}{{\ttfamily
  0705.0016}}].

\bibitem{Ryu:2006bv}
S.~Ryu and T.~Takayanagi, \emph{{Holographic derivation of entanglement entropy
  from AdS/CFT}},
  \href{https://doi.org/10.1103/PhysRevLett.96.181602}{\emph{Phys. Rev. Lett.}
  {\bfseries 96} (2006) 181602}
  [\href{https://arxiv.org/abs/hep-th/0603001}{{\ttfamily hep-th/0603001}}].

\bibitem{Chen:2018eqk}
B.~Chen, L.~Chen and P.-X.~Hao, \emph{{Entanglement entropy in
  $T\overline{T}$-deformed CFT}},
  \href{https://doi.org/10.1103/PhysRevD.98.086025}{\emph{Phys. Rev. D}
  {\bfseries 98} (2018) 086025}
  [\href{https://arxiv.org/abs/1807.08293}{{\ttfamily 1807.08293}}].

\bibitem{Jeong:2019ylz}
H.-S.~Jeong, K.-Y.~Kim and M.~Nishida, \emph{{Entanglement and R\'enyi entropy
  of multiple intervals in $T\overline{T}$-deformed CFT and holography}},
  \href{https://doi.org/10.1103/PhysRevD.100.106015}{\emph{Phys. Rev. D}
  {\bfseries 100} (2019) 106015}
  [\href{https://arxiv.org/abs/1906.03894}{{\ttfamily 1906.03894}}].

\bibitem{Jiang:2019epa}
Y.~Jiang, \emph{{A pedagogical review on solvable irrelevant deformations of 2D
  quantum field theory}},
  \href{https://doi.org/10.1088/1572-9494/abe4c9}{\emph{Commun. Theor. Phys.}
  {\bfseries 73} (2021) 057201}
  [\href{https://arxiv.org/abs/1904.13376}{{\ttfamily 1904.13376}}].

\bibitem{Allameh:2021moy}
K.~Allameh, A.F.~Astaneh and A.~Hassanzadeh, \emph{{Aspects of holographic
  entanglement entropy for $T\bar{T}$-deformed CFTs}},
  \href{https://doi.org/10.1016/j.physletb.2022.136914}{\emph{Phys. Lett. B}
  {\bfseries 826} (2022) 136914}
  [\href{https://arxiv.org/abs/2111.11338}{{\ttfamily 2111.11338}}].

\bibitem{Setare:2022qls}
M.R.~Setare and S.N.~Sajadi, \emph{{Holographic entanglement entropy in
  $T{\bar{T}}$-deformed CFTs}},
  \href{https://doi.org/10.1007/s10714-022-02971-y}{\emph{Gen. Rel. Grav.}
  {\bfseries 54} (2022) 85} [\href{https://arxiv.org/abs/2203.16445}{{\ttfamily
  2203.16445}}].

\bibitem{He:2023xnb}
M.~He and Y.~Sun, \emph{{Holographic entanglement entropy in
  $T\overline{T}$-deformed AdS$_3$}},
  \href{https://doi.org/10.1016/j.nuclphysb.2023.116190}{\emph{Nucl. Phys. B}
  {\bfseries 990} (2023) 116190}
  [\href{https://arxiv.org/abs/2301.04435}{{\ttfamily 2301.04435}}].

\bibitem{Apolo:2023ckr}
L.~Apolo, P.-X.~Hao, W.-X.~Lai and W.~Song, \emph{{Extremal surfaces in glue-on
  AdS/$ T\overline{T} $ holography}},
  \href{https://doi.org/10.1007/JHEP01(2024)054}{\emph{JHEP} {\bfseries 01}
  (2024) 054} [\href{https://arxiv.org/abs/2311.04883}{{\ttfamily
  2311.04883}}].

\bibitem{Apolo:2023vnm}
L.~Apolo, P.-X.~Hao, W.-X.~Lai and W.~Song, \emph{{Glue-on AdS holography for $
  T\overline{T} $-deformed CFTs}},
  \href{https://doi.org/10.1007/JHEP06(2023)117}{\emph{JHEP} {\bfseries 06}
  (2023) 117} [\href{https://arxiv.org/abs/2303.04836}{{\ttfamily
  2303.04836}}].

\bibitem{Dutta:2019gen}
S.~Dutta and T.~Faulkner, \emph{{A canonical purification for the entanglement
  wedge cross-section}},
  \href{https://doi.org/10.1007/JHEP03(2021)178}{\emph{JHEP} {\bfseries 03}
  (2021) 178} [\href{https://arxiv.org/abs/1905.00577}{{\ttfamily
  1905.00577}}].

\bibitem{Nakata:2020luh}
Y.~Nakata, T.~Takayanagi, Y.~Taki, K.~Tamaoka and Z.~Wei, \emph{{New
  holographic generalization of entanglement entropy}},
  \href{https://doi.org/10.1103/PhysRevD.103.026005}{\emph{Phys. Rev. D}
  {\bfseries 103} (2021) 026005}
  [\href{https://arxiv.org/abs/2005.13801}{{\ttfamily 2005.13801}}].

\bibitem{Basu:2024bal}
D.~Basu and V.~Raj, \emph{{Reflected entropy and timelike entanglement in $
  T\overline{T} $-deformed CFT2s}},
  \href{https://doi.org/10.1103/PhysRevD.110.046009}{\emph{Phys. Rev. D}
  {\bfseries 110} (2024) 046009}
  [\href{https://arxiv.org/abs/2402.07253}{{\ttfamily 2402.07253}}].

\bibitem{He:2023wko}
S.~He, J.~Yang, Y.-X.~Zhang and Z.-X.~Zhao, \emph{{Pseudo entropy of primary
  operators in $ T\overline{T}/J\overline{T} $-deformed CFTs}},
  \href{https://doi.org/10.1007/JHEP09(2023)025}{\emph{JHEP} {\bfseries 09}
  (2023) 025} [\href{https://arxiv.org/abs/2305.10984}{{\ttfamily
  2305.10984}}].

\bibitem{He:2019vzf}
S.~He and H.~Shu, \emph{{Correlation functions, entanglement and chaos in the $
  T\overline{T}/J\overline{T} $-deformed CFTs}},
  \href{https://doi.org/10.1007/JHEP02(2020)088}{\emph{JHEP} {\bfseries 02}
  (2020) 088} [\href{https://arxiv.org/abs/1907.12603}{{\ttfamily
  1907.12603}}].

\bibitem{Asrat:2020uib}
M.~Asrat and J.~Kudler-Flam, \emph{{$T\bar{T}$, the entanglement wedge cross
  section, and the breakdown of the split property}},
  \href{https://doi.org/10.1103/PhysRevD.102.045009}{\emph{Phys. Rev. D}
  {\bfseries 102} (2020) 045009}
  [\href{https://arxiv.org/abs/2005.08972}{{\ttfamily 2005.08972}}].

\bibitem{Khoeini-Moghaddam:2020ymm}
S.~Khoeini-Moghaddam, F.~Omidi and C.~Paul, \emph{{Aspects of Hyperscaling
  Violating Geometries at Finite Cutoff}},
  \href{https://doi.org/10.1007/JHEP02(2021)121}{\emph{JHEP} {\bfseries 02}
  (2021) 121} [\href{https://arxiv.org/abs/2011.00305}{{\ttfamily
  2011.00305}}].

\bibitem{Li:2020pwa}
Y.~Li and Y.~Zhou, \emph{{Cutoff AdS$_{3}$ versus $ T\overline{T} $ CFT$_{2}$
  in the large central charge sector: correlators of energy-momentum tensor}},
  \href{https://doi.org/10.1007/JHEP12(2020)168}{\emph{JHEP} {\bfseries 12}
  (2020) 168} [\href{https://arxiv.org/abs/2005.01693}{{\ttfamily
  2005.01693}}].

\bibitem{Jeong:2022jmp}
H.-S.~Jeong, W.-B.~Pan, Y.-W.~Sun and Y.-T.~Wang, \emph{{Holographic study of $
  T\overline{T} $ like deformed HV QFTs: holographic entanglement entropy}},
  \href{https://doi.org/10.1007/JHEP02(2023)018}{\emph{JHEP} {\bfseries 02}
  (2023) 018} [\href{https://arxiv.org/abs/2211.00518}{{\ttfamily
  2211.00518}}].

\bibitem{He:2023obo}
M.~He, J.~Hou and Y.~Jiang, \emph{{$ T\overline{T} $-deformed entanglement
  entropy for IQFT}},
  \href{https://doi.org/10.1007/JHEP03(2024)056}{\emph{JHEP} {\bfseries 03}
  (2024) 056} [\href{https://arxiv.org/abs/2306.07784}{{\ttfamily
  2306.07784}}].

\bibitem{Tian:2023fgf}
J.~Tian, \emph{{On-shell action of $\text{T}\bar{\text{T}}$-deformed
  Holographic CFTs}},  \href{https://arxiv.org/abs/2306.01258}{{\ttfamily
  2306.01258}}.

\bibitem{Basu:2023bov}
D.~Basu, Lavish and B.~Paul, \emph{{Entanglement negativity in
  $T\overline{T}$-deformed CFT2s}},
  \href{https://doi.org/10.1103/PhysRevD.107.126026}{\emph{Phys. Rev. D}
  {\bfseries 107} (2023) 126026}
  [\href{https://arxiv.org/abs/2302.11435}{{\ttfamily 2302.11435}}].

\bibitem{Basu:2023aqz}
D.~Basu, S.~Biswas, A.~Dey, B.~Paul and G.~Sengupta, \emph{{Odd entanglement
  entropy in $ T\overline{T}$ deformed CFT2s and holography}},
  \href{https://doi.org/10.1103/PhysRevD.108.126013}{\emph{Phys. Rev. D}
  {\bfseries 108} (2023) 126013}
  [\href{https://arxiv.org/abs/2307.04832}{{\ttfamily 2307.04832}}].

\bibitem{He:2024pbp}
M.~He, \emph{{One-loop partition functions in $T\overline{T }$-deformed
  AdS$_{3}$}}, \href{https://doi.org/10.1007/JHEP05(2024)067}{\emph{JHEP}
  {\bfseries 05} (2024) 067}
  [\href{https://arxiv.org/abs/2401.09879}{{\ttfamily 2401.09879}}].

\bibitem{Tsolakidis:2024wut}
E.~Tsolakidis, \emph{{Massive gravity generalization of $ T\overline{T} $
  deformations}}, \href{https://doi.org/10.1007/JHEP09(2024)167}{\emph{JHEP}
  {\bfseries 09} (2024) 167}
  [\href{https://arxiv.org/abs/2405.07967}{{\ttfamily 2405.07967}}].

\bibitem{Babaei-Aghbolagh:2024hti}
H.~Babaei-Aghbolagh, S.~He, T.~Morone, H.~Ouyang and R.~Tateo, \emph{{Geometric
  Formulation of Generalized Root-$ T\overline{T}$ Deformations}},
  \href{https://doi.org/10.1103/PhysRevLett.133.111602}{\emph{Phys. Rev. Lett.}
  {\bfseries 133} (2024) 111602}
  [\href{https://arxiv.org/abs/2405.03465}{{\ttfamily 2405.03465}}].

\bibitem{He:2025ppz}
S.~He, Y.~Li, H.~Ouyang and Y.~Sun, \emph{{$T\overline{T}$ Deformation:
  Introduction and Some Recent Advances}},
  \href{https://arxiv.org/abs/2503.09997}{{\ttfamily 2503.09997}}.

\bibitem{He:2025fdz}
M.~He, \emph{{Root-$T\overline{T}$ deformed CFT partition functions at large
  central charge}},  \href{https://arxiv.org/abs/2505.12093}{{\ttfamily
  2505.12093}}.

\bibitem{Witten:2001kn}
E.~Witten, \emph{{Quantum gravity in de Sitter space}},  in \emph{{Strings
  2001: International Conference}}, (Mumbai, India), 6, 2001
  [\href{https://arxiv.org/abs/hep-th/0106109}{{\ttfamily hep-th/0106109}}].

\bibitem{Maldacena:2002vr}
J.M.~Maldacena, \emph{{Non-Gaussian features of primordial fluctuations in
  single field inflationary models}},
  \href{https://doi.org/10.1088/1126-6708/2003/05/013}{\emph{JHEP} {\bfseries
  05} (2003) 013} [\href{https://arxiv.org/abs/astro-ph/0210603}{{\ttfamily
  astro-ph/0210603}}].

\bibitem{Anninos:2012ft}
D.~Anninos, F.~Denef and D.~Harlow, \emph{{Wave function of
  Vasiliev\textquoteright{}s universe: A few slices thereof}},
  \href{https://doi.org/10.1103/PhysRevD.88.084049}{\emph{Phys. Rev. D}
  {\bfseries 88} (2013) 084049}
  [\href{https://arxiv.org/abs/1207.5517}{{\ttfamily 1207.5517}}].

\bibitem{Miyaji:2015yva}
M.~Miyaji and T.~Takayanagi, \emph{{Surface/State Correspondence as a
  Generalized Holography}},
  \href{https://doi.org/10.1093/ptep/ptv089}{\emph{PTEP} {\bfseries 2015}
  (2015) 073B03} [\href{https://arxiv.org/abs/1503.03542}{{\ttfamily
  1503.03542}}].

\bibitem{Narayan:2017xca}
K.~Narayan, \emph{{On extremal surfaces and de Sitter entropy}},
  \href{https://doi.org/10.1016/j.physletb.2018.02.010}{\emph{Phys. Lett. B}
  {\bfseries 779} (2018) 214}
  [\href{https://arxiv.org/abs/1711.01107}{{\ttfamily 1711.01107}}].

\bibitem{Hikida:2021ese}
Y.~Hikida, T.~Nishioka, T.~Takayanagi and Y.~Taki, \emph{{Holography in de
  Sitter Space via Chern-Simons Gauge Theory}},
  \href{https://doi.org/10.1103/PhysRevLett.129.041601}{\emph{Phys. Rev. Lett.}
  {\bfseries 129} (2022) 041601}
  [\href{https://arxiv.org/abs/2110.03197}{{\ttfamily 2110.03197}}].

\bibitem{Chen:2022ozy}
H.-Y.~Chen and Y.~Hikida, \emph{{Three-Dimensional de Sitter Holography and
  Bulk Correlators at Late Time}},
  \href{https://doi.org/10.1103/PhysRevLett.129.061601}{\emph{Phys. Rev. Lett.}
  {\bfseries 129} (2022) 061601}
  [\href{https://arxiv.org/abs/2204.04871}{{\ttfamily 2204.04871}}].

\bibitem{Chang:2023gkt}
J.-C.~Chang, S.~He, Y.-X.~Liu and L.~Zhao, \emph{{Island formula in Planck
  brane}}, \href{https://doi.org/10.1007/JHEP11(2023)006}{\emph{JHEP}
  {\bfseries 11} (2023) 006}
  [\href{https://arxiv.org/abs/2308.03645}{{\ttfamily 2308.03645}}].

\bibitem{Chen:2023prz}
H.-Y.~Chen, Y.~Hikida, Y.~Taki and T.~Uetoko, \emph{{Complex saddles of
  three-dimensional de Sitter gravity via holography}},
  \href{https://doi.org/10.1103/PhysRevD.107.L101902}{\emph{Phys. Rev. D}
  {\bfseries 107} (2023) L101902}
  [\href{https://arxiv.org/abs/2302.09219}{{\ttfamily 2302.09219}}].

\bibitem{Doi:2024nty}
K.~Doi, N.~Ogawa, K.~Shinmyo, Y.-k.~Suzuki and T.~Takayanagi, \emph{{Probing de
  Sitter space using CFT states}},
  \href{https://doi.org/10.1007/JHEP02(2025)093}{\emph{JHEP} {\bfseries 02}
  (2025) 093} [\href{https://arxiv.org/abs/2405.14237}{{\ttfamily
  2405.14237}}].

\bibitem{Strominger:2001pn}
A.~Strominger, \emph{{The dS / CFT correspondence}},
  \href{https://doi.org/10.1088/1126-6708/2001/10/034}{\emph{JHEP} {\bfseries
  10} (2001) 034} [\href{https://arxiv.org/abs/hep-th/0106113}{{\ttfamily
  hep-th/0106113}}].

\bibitem{Strominger:2001gp}
A.~Strominger, \emph{{Inflation and the dS / CFT correspondence}},
  \href{https://doi.org/10.1088/1126-6708/2001/11/049}{\emph{JHEP} {\bfseries
  11} (2001) 049} [\href{https://arxiv.org/abs/hep-th/0110087}{{\ttfamily
  hep-th/0110087}}].

\bibitem{Anninos:2011ui}
D.~Anninos, T.~Hartman and A.~Strominger, \emph{{Higher Spin Realization of the
  dS/CFT Correspondence}},
  \href{https://doi.org/10.1088/1361-6382/34/1/015009}{\emph{Class. Quant.
  Grav.} {\bfseries 34} (2017) 015009}
  [\href{https://arxiv.org/abs/1108.5735}{{\ttfamily 1108.5735}}].

\bibitem{Ng:2012xp}
G.S.~Ng and A.~Strominger, \emph{{State/Operator Correspondence in Higher-Spin
  dS/CFT}}, \href{https://doi.org/10.1088/0264-9381/30/10/104002}{\emph{Class.
  Quant. Grav.} {\bfseries 30} (2013) 104002}
  [\href{https://arxiv.org/abs/1204.1057}{{\ttfamily 1204.1057}}].

\bibitem{Hikida:2022ltr}
Y.~Hikida, T.~Nishioka, T.~Takayanagi and Y.~Taki, \emph{{CFT duals of
  three-dimensional de Sitter gravity}},
  \href{https://doi.org/10.1007/JHEP05(2022)129}{\emph{JHEP} {\bfseries 05}
  (2022) 129} [\href{https://arxiv.org/abs/2203.02852}{{\ttfamily
  2203.02852}}].

\bibitem{Doi:2022iyj}
K.~Doi, J.~Harper, A.~Mollabashi, T.~Takayanagi and Y.~Taki,
  \emph{{Pseudoentropy in dS/CFT and Timelike Entanglement Entropy}},
  \href{https://doi.org/10.1103/PhysRevLett.130.031601}{\emph{Phys. Rev. Lett.}
  {\bfseries 130} (2023) 031601}
  [\href{https://arxiv.org/abs/2210.09457}{{\ttfamily 2210.09457}}].

\bibitem{Narayan:2015vda}
K.~Narayan, \emph{{Extremal surfaces in de Sitter spacetime}},
  \href{https://doi.org/10.1103/PhysRevD.91.126011}{\emph{Phys. Rev. D}
  {\bfseries 91} (2015) 126011}
  [\href{https://arxiv.org/abs/1501.03019}{{\ttfamily 1501.03019}}].

\bibitem{Narayan:2015oka}
K.~Narayan, \emph{{de Sitter space and extremal surfaces for spheres}},
  \href{https://doi.org/10.1016/j.physletb.2015.12.019}{\emph{Phys. Lett. B}
  {\bfseries 753} (2016) 308}
  [\href{https://arxiv.org/abs/1504.07430}{{\ttfamily 1504.07430}}].

\bibitem{Narayan:2016xwq}
K.~Narayan, \emph{{On $dS_4$ extremal surfaces and entanglement entropy in some
  ghost CFTs}}, \href{https://doi.org/10.1103/PhysRevD.94.046001}{\emph{Phys.
  Rev. D} {\bfseries 94} (2016) 046001}
  [\href{https://arxiv.org/abs/1602.06505}{{\ttfamily 1602.06505}}].

\bibitem{Narayan:2022afv}
K.~Narayan, \emph{{de Sitter space, extremal surfaces, and time entanglement}},
  \href{https://doi.org/10.1103/PhysRevD.107.126004}{\emph{Phys. Rev. D}
  {\bfseries 107} (2023) 126004}
  [\href{https://arxiv.org/abs/2210.12963}{{\ttfamily 2210.12963}}].

\bibitem{Narayan:2023zen}
K.~Narayan, \emph{{Further remarks on de Sitter space, extremal surfaces, and
  time entanglement}},
  \href{https://doi.org/10.1103/PhysRevD.109.086009}{\emph{Phys. Rev. D}
  {\bfseries 109} (2024) 086009}
  [\href{https://arxiv.org/abs/2310.00320}{{\ttfamily 2310.00320}}].

\bibitem{Sato:2015tta}
Y.~Sato, \emph{{Comments on Entanglement Entropy in the dS/CFT
  Correspondence}},
  \href{https://doi.org/10.1103/PhysRevD.91.086009}{\emph{Phys. Rev. D}
  {\bfseries 91} (2015) 086009}
  [\href{https://arxiv.org/abs/1501.04903}{{\ttfamily 1501.04903}}].

\bibitem{Alishahiha:2004md}
M.~Alishahiha, A.~Karch, E.~Silverstein and D.~Tong, \emph{{The dS/dS
  correspondence}}, \href{https://doi.org/10.1063/1.1848341}{\emph{AIP Conf.
  Proc.} {\bfseries 743} (2004) 393}
  [\href{https://arxiv.org/abs/hep-th/0407125}{{\ttfamily hep-th/0407125}}].

\bibitem{Alishahiha:2005dj}
M.~Alishahiha, A.~Karch and E.~Silverstein, \emph{{Hologravity}},
  \href{https://doi.org/10.1088/1126-6708/2005/06/028}{\emph{JHEP} {\bfseries
  06} (2005) 028} [\href{https://arxiv.org/abs/hep-th/0504056}{{\ttfamily
  hep-th/0504056}}].

\bibitem{Dong:2018cuv}
X.~Dong, E.~Silverstein and G.~Torroba, \emph{{De Sitter Holography and
  Entanglement Entropy}},
  \href{https://doi.org/10.1007/JHEP07(2018)050}{\emph{JHEP} {\bfseries 07}
  (2018) 050} [\href{https://arxiv.org/abs/1804.08623}{{\ttfamily
  1804.08623}}].

\bibitem{Geng:2019bnn}
H.~Geng, S.~Grieninger and A.~Karch, \emph{{Entropy, Entanglement and Swampland
  Bounds in DS/dS}}, \href{https://doi.org/10.1007/JHEP06(2019)105}{\emph{JHEP}
  {\bfseries 06} (2019) 105}
  [\href{https://arxiv.org/abs/1904.02170}{{\ttfamily 1904.02170}}].

\bibitem{Geng:2019ruz}
H.~Geng, \emph{{Some Information Theoretic Aspects of De-Sitter Holography}},
  \href{https://doi.org/10.1007/JHEP02(2020)005}{\emph{JHEP} {\bfseries 02}
  (2020) 005} [\href{https://arxiv.org/abs/1911.02644}{{\ttfamily
  1911.02644}}].

\bibitem{Geng:2020kxh}
H.~Geng, \emph{{Non-local entanglement and fast scrambling in de-Sitter
  holography}}, \href{https://doi.org/10.1016/j.aop.2021.168402}{\emph{Annals
  Phys.} {\bfseries 426} (2021) 168402}
  [\href{https://arxiv.org/abs/2005.00021}{{\ttfamily 2005.00021}}].

\bibitem{Geng:2021wcq}
H.~Geng, Y.~Nomura and H.-Y.~Sun, \emph{{Information paradox and its resolution
  in de Sitter holography}},
  \href{https://doi.org/10.1103/PhysRevD.103.126004}{\emph{Phys. Rev. D}
  {\bfseries 103} (2021) 126004}
  [\href{https://arxiv.org/abs/2103.07477}{{\ttfamily 2103.07477}}].

\bibitem{Kawamoto:2023nki}
T.~Kawamoto, S.-M.~Ruan, Y.-k.~Suzuki and T.~Takayanagi, \emph{{A half de
  Sitter holography}},
  \href{https://doi.org/10.1007/JHEP10(2023)137}{\emph{JHEP} {\bfseries 10}
  (2023) 137} [\href{https://arxiv.org/abs/2306.07575}{{\ttfamily
  2306.07575}}].

\bibitem{Chen:2023eic}
D.~Chen, X.~Jiang and H.~Yang, \emph{{Holographic $T\overline{T}$ deformed
  entanglement entropy in dS$_3$/CFT$_2$}},
  \href{https://doi.org/10.1103/PhysRevD.109.026011}{\emph{Phys. Rev. D}
  {\bfseries 109} (2024) 026011}
  [\href{https://arxiv.org/abs/2307.04673}{{\ttfamily 2307.04673}}].

\bibitem{Araujo-Regado:2022gvw}
G.~Araujo-Regado, R.~Khan and A.C.~Wall, \emph{{Cauchy slice holography: a new
  AdS/CFT dictionary}},
  \href{https://doi.org/10.1007/JHEP03(2023)026}{\emph{JHEP} {\bfseries 03}
  (2023) 026} [\href{https://arxiv.org/abs/2204.00591}{{\ttfamily
  2204.00591}}].

\bibitem{Araujo-Regado:2022jpj}
G.~Araujo-Regado, \emph{{Holographic Cosmology on Closed Slices in 2+1
  Dimensions}},  \href{https://arxiv.org/abs/2212.03219}{{\ttfamily
  2212.03219}}.

\bibitem{Gorbenko:2018oov}
V.~Gorbenko, E.~Silverstein and G.~Torroba, \emph{{dS/dS and $ T\overline{T}
  $}}, \href{https://doi.org/10.1007/JHEP03(2019)085}{\emph{JHEP} {\bfseries
  03} (2019) 085} [\href{https://arxiv.org/abs/1811.07965}{{\ttfamily
  1811.07965}}].

\bibitem{Shyam:2021ciy}
V.~Shyam, \emph{{$ \mathrm{T}\overline{\mathrm{T}} $ +
  \ensuremath{\Lambda}$_{2}$ deformed CFT on the stretched dS$_{3}$ horizon}},
  \href{https://doi.org/10.1007/JHEP04(2022)052}{\emph{JHEP} {\bfseries 04}
  (2022) 052} [\href{https://arxiv.org/abs/2106.10227}{{\ttfamily
  2106.10227}}].

\bibitem{Coleman:2021nor}
E.~Coleman, E.A.~Mazenc, V.~Shyam, E.~Silverstein, R.M.~Soni, G.~Torroba
  et~al., \emph{{De Sitter microstates from T$ \overline{T} $ +
  \ensuremath{\Lambda}$_{2}$ and the Hawking-Page transition}},
  \href{https://doi.org/10.1007/JHEP07(2022)140}{\emph{JHEP} {\bfseries 07}
  (2022) 140} [\href{https://arxiv.org/abs/2110.14670}{{\ttfamily
  2110.14670}}].

\bibitem{Torroba:2022jrk}
G.~Torroba, \emph{{$ T\overline{T} $ + \ensuremath{\Lambda}$_{2}$ from a 2d
  gravity path integral}},
  \href{https://doi.org/10.1007/JHEP01(2023)163}{\emph{JHEP} {\bfseries 01}
  (2023) 163} [\href{https://arxiv.org/abs/2212.04512}{{\ttfamily
  2212.04512}}].

\bibitem{Batra:2024kjl}
G.~Batra, G.B.~De~Luca, E.~Silverstein, G.~Torroba and S.~Yang,
  \emph{{Bulk-local dS$_{3}$ holography: the matter with $ T\overline{T} $ +
  \ensuremath{\Lambda}$_{2}$}},
  \href{https://doi.org/10.1007/JHEP10(2024)072}{\emph{JHEP} {\bfseries 10}
  (2024) 072} [\href{https://arxiv.org/abs/2403.01040}{{\ttfamily
  2403.01040}}].

\bibitem{Calabrese:2009qy}
P.~Calabrese and J.~Cardy, \emph{{Entanglement entropy and conformal field
  theory}}, \href{https://doi.org/10.1088/1751-8113/42/50/504005}{\emph{J.
  Phys. A} {\bfseries 42} (2009) 504005}
  [\href{https://arxiv.org/abs/0905.4013}{{\ttfamily 0905.4013}}].

\bibitem{Donnelly:2018bef}
W.~Donnelly and V.~Shyam, \emph{{Entanglement entropy and $T \overline{T}$
  deformation}},
  \href{https://doi.org/10.1103/PhysRevLett.121.131602}{\emph{Phys. Rev. Lett.}
  {\bfseries 121} (2018) 131602}
  [\href{https://arxiv.org/abs/1806.07444}{{\ttfamily 1806.07444}}].

\bibitem{Caputa:2019pam}
P.~Caputa, S.~Datta and V.~Shyam, \emph{{Sphere partition functions
  \textbackslash{}\& cut-off AdS}},
  \href{https://doi.org/10.1007/JHEP05(2019)112}{\emph{JHEP} {\bfseries 05}
  (2019) 112} [\href{https://arxiv.org/abs/1902.10893}{{\ttfamily
  1902.10893}}].

\bibitem{Deng:2023pjs}
F.~Deng, Z.~Wang and Y.~Zhou, \emph{{End of the world brane meets $
  T\overline{T} $}}, \href{https://doi.org/10.1007/JHEP07(2024)036}{\emph{JHEP}
  {\bfseries 07} (2024) 036}
  [\href{https://arxiv.org/abs/2310.15031}{{\ttfamily 2310.15031}}].

\bibitem{Grieninger:2019zts}
S.~Grieninger, \emph{{Entanglement entropy and $ T\overline{T} $ deformations
  beyond antipodal points from holography}},
  \href{https://doi.org/10.1007/JHEP11(2019)171}{\emph{JHEP} {\bfseries 11}
  (2019) 171} [\href{https://arxiv.org/abs/1908.10372}{{\ttfamily
  1908.10372}}].

\bibitem{Wilson:1973jj}
K.G.~Wilson and J.B.~Kogut, \emph{{The Renormalization group and the epsilon
  expansion}}, \href{https://doi.org/10.1016/0370-1573(74)90023-4}{\emph{Phys.
  Rept.} {\bfseries 12} (1974) 75}.

\bibitem{Polchinski:1983gv}
J.~Polchinski, \emph{{Renormalization and Effective Lagrangians}},
  \href{https://doi.org/10.1016/0550-3213(84)90287-6}{\emph{Nucl. Phys. B}
  {\bfseries 231} (1984) 269}.

\bibitem{Lewkowycz:2013nqa}
A.~Lewkowycz and J.~Maldacena, \emph{{Generalized gravitational entropy}},
  \href{https://doi.org/10.1007/JHEP08(2013)090}{\emph{JHEP} {\bfseries 08}
  (2013) 090} [\href{https://arxiv.org/abs/1304.4926}{{\ttfamily 1304.4926}}].

\bibitem{Lewkowycz:2019xse}
A.~Lewkowycz, J.~Liu, E.~Silverstein and G.~Torroba, \emph{{$ T\overline{T} $
  and EE, with implications for (A)dS subregion encodings}},
  \href{https://doi.org/10.1007/JHEP04(2020)152}{\emph{JHEP} {\bfseries 04}
  (2020) 152} [\href{https://arxiv.org/abs/1909.13808}{{\ttfamily
  1909.13808}}].

\bibitem{Casini:2011kv}
H.~Casini, M.~Huerta and R.C.~Myers, \emph{{Towards a derivation of holographic
  entanglement entropy}},
  \href{https://doi.org/10.1007/JHEP05(2011)036}{\emph{JHEP} {\bfseries 05}
  (2011) 036} [\href{https://arxiv.org/abs/1102.0440}{{\ttfamily 1102.0440}}].

\bibitem{Brown1986}
J.D.~Brown and M.~Henneaux, \emph{{Central charges in the canonical realization
  of asymptotic symmetries: An example from three dimensional gravity}},
  \href{https://doi.org/10.1007/BF01211590}{\emph{Commun. Math. Phys.}
  {\bfseries 104} (1986) 207}.

\bibitem{Shyam:2017znq}
V.~Shyam, \emph{{Background independent holographic dual to $T\bar{T}$ deformed
  CFT with large central charge in 2 dimensions}},
  \href{https://doi.org/10.1007/JHEP10(2017)108}{\emph{JHEP} {\bfseries 10}
  (2017) 108} [\href{https://arxiv.org/abs/1707.08118}{{\ttfamily
  1707.08118}}].

\bibitem{Hartman:2018tkw}
T.~Hartman, J.~Kruthoff, E.~Shaghoulian and A.~Tajdini, \emph{{Holography at
  finite cutoff with a $T^2$ deformation}},
  \href{https://doi.org/10.1007/JHEP03(2019)004}{\emph{JHEP} {\bfseries 03}
  (2019) 004} [\href{https://arxiv.org/abs/1807.11401}{{\ttfamily
  1807.11401}}].

\bibitem{Balasubramanian:1999re}
V.~Balasubramanian and P.~Kraus, \emph{{A Stress tensor for Anti-de Sitter
  gravity}}, \href{https://doi.org/10.1007/s002200050764}{\emph{Commun. Math.
  Phys.} {\bfseries 208} (1999) 413}
  [\href{https://arxiv.org/abs/hep-th/9902121}{{\ttfamily hep-th/9902121}}].

\bibitem{FeffermanGraham1985}
C.~Fefferman and C.R.~Graham, \emph{Conformal invariants},  in \emph{\'Elie
  Cartan et les math\'ematiques d'aujourd'hui -- Lyon, 25--29 June 1984},
  no.~S131 in Ast\'erisque, (Lyon, France), pp.~95--116, Soci\'et\'e
  Math\'ematique de France (1985).

\bibitem{Arnowitt:1962hi}
R.L.~Arnowitt, S.~Deser and C.W.~Misner, \emph{{The Dynamics of general
  relativity}}, \href{https://doi.org/10.1007/s10714-008-0661-1}{\emph{Gen.
  Rel. Grav.} {\bfseries 40} (2008) 1997}
  [\href{https://arxiv.org/abs/gr-qc/0405109}{{\ttfamily gr-qc/0405109}}].

\bibitem{Witten:2022xxp}
E.~Witten, \emph{{A note on the canonical formalism for gravity}},
  \href{https://doi.org/10.4310/ATMP.2023.v27.n1.a6}{\emph{Adv. Theor. Math.
  Phys.} {\bfseries 27} (2023) 311}
  [\href{https://arxiv.org/abs/2212.08270}{{\ttfamily 2212.08270}}].

\bibitem{DeWitt:1967yk}
B.S.~DeWitt, \emph{{Quantum Theory of Gravity. 1. The Canonical Theory}},
  \href{https://doi.org/10.1103/PhysRev.160.1113}{\emph{Phys. Rev.} {\bfseries
  160} (1967) 1113}.

\bibitem{Chakraborty:2023yed}
T.~Chakraborty, J.~Chakravarty, V.~Godet, P.~Paul and S.~Raju, \emph{{The
  Hilbert space of de Sitter quantum gravity}},
  \href{https://doi.org/10.1007/JHEP01(2024)132}{\emph{JHEP} {\bfseries 01}
  (2024) 132} [\href{https://arxiv.org/abs/2303.16315}{{\ttfamily
  2303.16315}}].

\bibitem{Cianfrani:2013oja}
F.~Cianfrani and J.~Kowalski-Glikman, \emph{{Wheeler-DeWitt equation and
  AdS/CFT correspondence}},
  \href{https://doi.org/10.1016/j.physletb.2013.07.034}{\emph{Phys. Lett. B}
  {\bfseries 725} (2013) 463}
  [\href{https://arxiv.org/abs/1306.0353}{{\ttfamily 1306.0353}}].

\bibitem{Chakraborty:2023los}
T.~Chakraborty, J.~Chakravarty, V.~Godet, P.~Paul and S.~Raju,
  \emph{{Holography of information in de Sitter space}},
  \href{https://doi.org/10.1007/JHEP12(2023)120}{\emph{JHEP} {\bfseries 12}
  (2023) 120} [\href{https://arxiv.org/abs/2303.16316}{{\ttfamily
  2303.16316}}].

\bibitem{Dong:2023bax}
X.~Dong, G.N.~Remmen, D.~Wang, W.W.~Weng and C.-H.~Wu, \emph{{Holographic
  entanglement from the UV to the IR}},
  \href{https://doi.org/10.1007/JHEP11(2023)207}{\emph{JHEP} {\bfseries 11}
  (2023) 207} [\href{https://arxiv.org/abs/2308.07952}{{\ttfamily
  2308.07952}}].

\bibitem{Headrick:2010zt}
M.~Headrick, \emph{{Entanglement Renyi entropies in holographic theories}},
  \href{https://doi.org/10.1103/PhysRevD.82.126010}{\emph{Phys. Rev. D}
  {\bfseries 82} (2010) 126010}
  [\href{https://arxiv.org/abs/1006.0047}{{\ttfamily 1006.0047}}].

\bibitem{Hung:2011nu}
L.-Y.~Hung, R.C.~Myers, M.~Smolkin and A.~Yale, \emph{{Holographic Calculations
  of Renyi Entropy}},
  \href{https://doi.org/10.1007/JHEP12(2011)047}{\emph{JHEP} {\bfseries 12}
  (2011) 047} [\href{https://arxiv.org/abs/1110.1084}{{\ttfamily 1110.1084}}].

\bibitem{Casini:2012ei}
H.~Casini and M.~Huerta, \emph{{On the RG running of the entanglement entropy
  of a circle}}, \href{https://doi.org/10.1103/PhysRevD.85.125016}{\emph{Phys.
  Rev. D} {\bfseries 85} (2012) 125016}
  [\href{https://arxiv.org/abs/1202.5650}{{\ttfamily 1202.5650}}].

\bibitem{Casini:2004bw}
H.~Casini and M.~Huerta, \emph{{A Finite entanglement entropy and the
  c-theorem}},
  \href{https://doi.org/10.1016/j.physletb.2004.08.072}{\emph{Phys. Lett. B}
  {\bfseries 600} (2004) 142}
  [\href{https://arxiv.org/abs/hep-th/0405111}{{\ttfamily hep-th/0405111}}].

\bibitem{Barnes:2004jj}
E.~Barnes, K.A.~Intriligator, B.~Wecht and J.~Wright, \emph{{Evidence for the
  strongest version of the 4d a-theorem, via a-maximization along RG flows}},
  \href{https://doi.org/10.1016/j.nuclphysb.2004.09.016}{\emph{Nucl. Phys. B}
  {\bfseries 702} (2004) 131}
  [\href{https://arxiv.org/abs/hep-th/0408156}{{\ttfamily hep-th/0408156}}].

\bibitem{Gukov:2015qea}
S.~Gukov, \emph{{Counting RG flows}},
  \href{https://doi.org/10.1007/JHEP01(2016)020}{\emph{JHEP} {\bfseries 01}
  (2016) 020} [\href{https://arxiv.org/abs/1503.01474}{{\ttfamily
  1503.01474}}].

\bibitem{Zamolodchikov:1986gt}
A.B.~Zamolodchikov, \emph{{Irreversibility of the Flux of the Renormalization
  Group in a 2D Field Theory}}, {\emph{JETP Lett.} {\bfseries 43} (1986) 730}.

\bibitem{deBoer:1999tgo}
J.~de~Boer, E.P.~Verlinde and H.L.~Verlinde, \emph{{On the holographic
  renormalization group}},
  \href{https://doi.org/10.1088/1126-6708/2000/08/003}{\emph{JHEP} {\bfseries
  08} (2000) 003} [\href{https://arxiv.org/abs/hep-th/9912012}{{\ttfamily
  hep-th/9912012}}].

\bibitem{Verlinde:1999xm}
E.P.~Verlinde and H.L.~Verlinde, \emph{{RG flow, gravity and the cosmological
  constant}}, \href{https://doi.org/10.1088/1126-6708/2000/05/034}{\emph{JHEP}
  {\bfseries 05} (2000) 034}
  [\href{https://arxiv.org/abs/hep-th/9912018}{{\ttfamily hep-th/9912018}}].

\bibitem{Heemskerk:2010hk}
I.~Heemskerk and J.~Polchinski, \emph{{Holographic and Wilsonian
  Renormalization Groups}},
  \href{https://doi.org/10.1007/JHEP06(2011)031}{\emph{JHEP} {\bfseries 06}
  (2011) 031} [\href{https://arxiv.org/abs/1010.1264}{{\ttfamily 1010.1264}}].

\bibitem{Nishioka:2018khk}
T.~Nishioka, \emph{{Entanglement entropy: holography and renormalization
  group}}, \href{https://doi.org/10.1103/RevModPhys.90.035007}{\emph{Rev. Mod.
  Phys.} {\bfseries 90} (2018) 035007}
  [\href{https://arxiv.org/abs/1801.10352}{{\ttfamily 1801.10352}}].

\bibitem{Grieninger:2023knz}
S.~Grieninger, K.~Ikeda and D.E.~Kharzeev, \emph{{Temporal entanglement entropy
  as a probe of renormalization group flow}},
  \href{https://doi.org/10.1007/JHEP05(2024)030}{\emph{JHEP} {\bfseries 05}
  (2024) 030} [\href{https://arxiv.org/abs/2312.08534}{{\ttfamily
  2312.08534}}].

\bibitem{Kamei:2024iei}
H.~Kamei, \emph{{Notes on the Exact RG equation and the Wheeler-DeWitt
  equation}},  \href{https://arxiv.org/abs/2402.16100}{{\ttfamily 2402.16100}}.

\bibitem{Nakayama:2013is}
Y.~Nakayama, \emph{{Scale invariance vs conformal invariance}},
  \href{https://doi.org/10.1016/j.physrep.2014.12.003}{\emph{Phys. Rept.}
  {\bfseries 569} (2015) 1} [\href{https://arxiv.org/abs/1302.0884}{{\ttfamily
  1302.0884}}].

\bibitem{Doi:2023zaf}
K.~Doi, J.~Harper, A.~Mollabashi, T.~Takayanagi and Y.~Taki, \emph{{Timelike
  entanglement entropy}},
  \href{https://doi.org/10.1007/JHEP05(2023)052}{\emph{JHEP} {\bfseries 05}
  (2023) 052} [\href{https://arxiv.org/abs/2302.11695}{{\ttfamily
  2302.11695}}].

\bibitem{Banados:1992wn}
M.~Banados, C.~Teitelboim and J.~Zanelli, \emph{{The Black hole in
  three-dimensional space-time}},
  \href{https://doi.org/10.1103/PhysRevLett.69.1849}{\emph{Phys. Rev. Lett.}
  {\bfseries 69} (1992) 1849}
  [\href{https://arxiv.org/abs/hep-th/9204099}{{\ttfamily hep-th/9204099}}].

\bibitem{Aharony:2023dod}
O.~Aharony and N.~Barel, \emph{{Correlation functions in $
  \textrm{T}\overline{\textrm{T}} $-deformed Conformal Field Theories}},
  \href{https://doi.org/10.1007/JHEP08(2023)035}{\emph{JHEP} {\bfseries 08}
  (2023) 035} [\href{https://arxiv.org/abs/2304.14091}{{\ttfamily
  2304.14091}}].

\bibitem{Witten:2018zxz}
E.~Witten, \emph{{APS Medal for Exceptional Achievement in Research: Invited
  article on entanglement properties of quantum field theory}},
  \href{https://doi.org/10.1103/RevModPhys.90.045003}{\emph{Rev. Mod. Phys.}
  {\bfseries 90} (2018) 045003}
  [\href{https://arxiv.org/abs/1803.04993}{{\ttfamily 1803.04993}}].

\end{thebibliography}\endgroup
\end{document}